\documentclass{aa}  
\usepackage[colorlinks=true,     linkcolor=blue, citecolor=blue, filecolor=blue, urlcolor=blue]{hyperref}
\usepackage{graphicx}
\usepackage{float}
\usepackage{txfonts}
\newcommand{\zh}{$\rm [Z/H]$}

\newcommand{\feh}{$\rm [Fe/H]$}

\newcommand{\mgfe}{$\rm [Mg/Fe]$}
\newcommand{\ofe}{$\rm [O/Fe]$}
\newcommand{\xfe}{$\rm [X/Fe]$}
\newcommand{\xh}{$\rm [X/H]$}
\newcommand{\nafe}{$\rm [Na/Fe]$}

\newcommand{\tife}{$\rm [Ti/Fe]$}

\newcommand{\tioi}{$\rm TiO1$}

\newcommand{\tioiir}{$\rm TiO2_{SDSS}$}
\newcommand{\tioii}{$\rm TiO2$}

\newcommand{\cafr}{$\rm Ca4227r$}
\newcommand{\cai}{$\rm Ca1$}
\newcommand{\caii}{$\rm Ca2$}

\newcommand{\hb}{$\rm H\beta$}

\newcommand{\hgf}{$\rm H\gamma_F$}

\newcommand{\hdf}{$\rm H\delta_F$}
\newcommand{\mgb}{$\rm Mgb5177$}
\newcommand{\nai}{$\rm NaD$}
\newcommand{\naii}{$\rm NaI8190$}

\newcommand{\nad}{$\rm NaD$}
\newcommand{\cnii}{$\rm CN2$}

\newcommand{\afe}{$\rm [\alpha/Fe]$}
\newcommand{\kms}{\,km\,s$^{-1}$}
\newcommand{\vrad}{$\rm V_{rad}$}

\newcommand{\age}{$\rm Age$}

\newcommand{\gammab}{$\rm \Gamma_b$}

\newcommand{\mgf}{$\rm Mg4780$}

\newcommand{\Age}{Age}
\newcommand{\mlr}{M/L$_r$}

\begin{document}
\titlerunning{Different paths, same appearance: an intermediate-age red nugget at $z\sim0.13$}
\authorrunning{F. La Barbera et al.}

   \title{Different paths, same appearance: an intermediate-age red nugget at $\rm z \sim 0.13$ residing in a low-mass group environment}

   \subtitle{}

   \author{F. La Barbera~\inst{1}, 
   F. Buitrago~\inst{2}, 
   I. Ferreras~\inst{3,4,5},
   C. Tortora~\inst{1}
          }

   \institute{INAF-Osservatorio Astronomico di Capodimonte, sal. Moiariello 16, Napoli, 80131, Italy\\
              \email{francesco.labarbera@inaf.it}
         \and
         Instituto de Astrof\'{\i}sica e Ci\^{e}ncias do Espa\c{c}o, Universidade de Lisboa, OAL, Tapada da Ajuda, PT-1349-018 Lisbon, Portugal
         \and
         Instituto de Astrof\'\i sica de Canarias, Calle V\'\i a L\'actea s/n, E-38205
         \and
         Departamento de Astrof\'\i sica, Universidad de La Laguna (ULL), E-38206  La Laguna, Tenerife, Spain
         \and
         Department of Physics and Astronomy, University College London, London WC1E 6BT, UK
   }

   \date{Received ; accepted}

   \abstract{
   { Massive ultra-compact galaxies} are commonly regarded as nearby relics of the compact quiescent population at $z\sim2$--3, and detailed studies have focused so far only on old systems in dense environments. Here we present a spatially resolved analysis of G\,79071, an ultra-compact ($R_{\rm e}<2$\,kpc), massive ($M_\star\!\sim\!10^{11}\,M_\odot$) { early-type galaxy (ETG)} at $z\sim0.13$ residing in a { low-mass group} environment, based on deep VLT/X-Shooter long-slit spectroscopy.
   We extract stellar kinematics out to $\sim4\,R_{\rm e}$ and constrain stellar population properties and the low-mass end of the stellar initial mass function (IMF), by combining full spectral fitting, full-index fitting, and index fitting, using different stellar population models. { G\,79071 shows significant rotation, consistent with a fast-rotator-like kinematic structure,} and is dominated by an intermediate-age stellar population ($\sim3$--4\,Gyr) with a flat age profile and no evidence for a significantly old ($\gtrsim5$\,Gyr) component. The central metallicity is supersolar and decreases with radius with a gradient typical of ETGs, while most abundance ratios show flat radial trends and are consistent with typical massive, low-redshift ETGs at similar velocity dispersion.  Moreover, metallicity and \nafe\ abundance are significantly enhanced,  reflecting a very efficient  chemical enrichment process.
{ Our analysis suggests that the IMF} remains bottom-heavy out to $\sim2\,R_{\rm e}$, implying a mass-excess factor $\alpha \gtrsim 2$. Jeans anisotropic models with an NFW halo are consistent with the IMF-based stellar mass normalization and with compactness-corrected (non-homologous) dynamical mass estimators, yielding
 a modest projected dark-matter fraction within one effective { radius, $f_{\rm DM}(<R_{\rm e})=0.18 \pm 0.07$}. These results show that { at least some massive compact galaxies} with a bottom-heavy IMF { can form} and persist at lower redshift and outside high-density environments, providing new constraints on the diversity of formation pathways for massive galaxies.
   }


   \maketitle
   \nolinenumbers
%

\section{Introduction}
\label{sec:intro}

Galaxy evolution unfolds over several gigayears, on timescales comparable to the age of the Universe. A direct
route to constrain this process is to identify systems whose mass assembly and star formation have largely ceased,
thereby preserving the physical conditions imprinted at early epochs. In this context, the most massive galaxies
($\rm M_\star \gtrsim 10^{11}\,M_\odot$) have attracted particular attention: they are expected to form preferentially in
rare, high-$\sigma$ peaks of the primordial density field and to undergo an ``accelerated'' evolution relative to the
general population. A striking manifestation of this channel is the presence at $\rm z\sim$2--3 of compact, quiescent
massive galaxies -- often displaying disky-elliptical morphologies--commonly referred to as ``red nuggets''
(e.g.~\citealt{Buitrago:2008,Damjanov:2009, Ferreras:2012, Buitrago:2018, HuertasCompany:2016, Toft:2017}).

For some time, the nearby Universe appeared largely devoid of
analogous compact quiescent systems.  However, deep, wide-area imaging
coupled to highly complete spectroscopy has uncovered a rare local
population of massive compact galaxies, including ``relic'' candidates
defined by extreme compactness (${\rm R_e}\lesssim2\,{\rm kpc}$) and
very old stellar populations ($\gtrsim 8$--$10\,{\rm Gyr}$;
\citealt{Trujillo:2014, FerreMateu:2017}).  These systems are key
laboratories for testing the physics of early compaction, quenching,
and black-hole--galaxy co-evolution, given their reported extreme
properties, such as \"{u}bermassive black holes \citep{Yildirim:17}, a
non-standard bottom-heavy stellar initial mass function~(IMF;
\citealt{NMN:15c}), and atypical globular-cluster
systems~\citep{Beasley:2018}.  Their diagnostic power is tempered by
their scarcity: typical number densities are of order $\sim
10^{-6}\,{\rm Mpc}^{-3} $, implying that one must search $\sim10^3$
massive galaxies at $\rm z \sim 0$ to find a single compact relic
candidate \citep{QuilisTrujillo:2013}.  { In the following, we use the term 
`red nugget' in a broader structural sense, to refer to massive,
compact, quiescent galaxies, while reserving the term ``relic'' for  systems that also host uniformly old stellar
populations.}

A further open issue concerns environment. Cosmological simulations often predict that relic systems should be
preferentially found in overdensities, where high velocity dispersions can suppress mergers and help preserve compact
remnants \citep[e.g.,][]{Stringer:2015}. Observational searches, however, suggest a more nuanced picture. In particular,
a systematic study in the Galaxy And Mass Assembly survey 
\citep[GAMA, ][]{Driver:11,Liske:15}, exploiting
the overlap with deep KiDS and VIKING imaging, identified a sample of massive compact galaxies at
$0.02<z<0.3$, many of which reside in small groups rather than rich clusters~\citep{Buitrago:2018}. This tension
raises a basic question: are massive compact galaxies a single evolutionary class whose survival depends primarily on
environment, or a heterogeneous population produced by multiple formation channels?

Progress requires moving beyond global, aperture-limited constraints. Spatially resolved stellar population studies
provide direct access to the radial imprint of the physical processes responsible for compaction and quenching (e.g.
inside-out versus outside-in formation, centrally concentrated starbursts, dissipative events, and the subsequent
efficiency of mixing). Despite their importance, only a handful of spatially resolved analyses exist for massive compact
early-type galaxies \citep[e.g.,][]{NMN:15c,FerreMateu:2017}, and most work has focused on
nearby, old ``relic'' benchmarks such as NGC\,1277. At the same time, massive compact galaxies span a range of stellar
population ages \citep[e.g.,][]{AFM:12} and environments~\citep{Tortora:2020, Scognamiglio:24}, suggesting that studying 
different subclasses of these systems may provide important insight into the formation pathways of the high--$z$ red--nugget population.
A step in this direction has recently been taken by the INSPIRE project~\citep{Spiniello:24}.
However, the low comoving number density of compact massive galaxies limits current samples and makes it difficult
to isolate the physical origin of compactness. As a first step, it is therefore essential to study in detail the
few available objects across a broad range of properties, most notably environment.

In this paper we present a detailed, spatially resolved study of GAMA
ID 79071 (hereafter G\,79071; RA = 14:39:51.26, DEC = +00:06:45.15
[J2000]), an ultra--compact (${\rm R_e}<2\,{\rm kpc}$), massive
($M_\star \sim 10^{11}\,M_\odot$) early-type galaxy at $z\sim0.13$,
residing in { a low-mass group of galaxies, with an estimated group
velocity dispersion} {$\sigma_{\rm group}$ = 63 $\pm$ 31 km s$^{-1}$
  \citep{Robotham:11}}.   G\,79071 is
an especially informative test case: if compact massive systems can
persist (or form) outside rich clusters, their kinematics, stellar
population gradients, chemical enrichment patterns, and IMF
constraints can be used to discriminate between competing scenarios,
including early dissipative compaction followed by long-term passive
evolution versus later formation or rejuvenation channels that can
also yield compact remnants.  By placing G\,79071 in the context of
the emerging, environmentally diverse population of massive compact
galaxies~\citep[e.g.,][]{Buitrago:2018, Tortora:2020,
  Scognamiglio:24}, we aim to clarify what these systems can reveal
about the origin of red nuggets and the physical conditions that allow
compactness to be established and maintained.
{ Throughout the manuscript, we refer to G\,79071 as an
``intermediate-age red nugget', emphasizing that it satisfies the
structural criteria of massive compact quiescent galaxies, while, as
shown below, it does not qualify as a classical (old) relic galaxy.}

This  paper  is  structured  as follows.  In  Sect.~\ref{sec:data}  we
describe the  X-Shooter observations  of G\,79071 and  data reduction.
Sect.~\ref{sec:models} describes the stellar population models adopted
in  this   work,  while   the  analysis   methods  are   presented  in
Sect.~\ref{sec:methods}.     Our     results      are     given     in
Sect.~\ref{sec:results},        including        the        kinematics
(Sect.~\ref{sec:kin}), the spatially resolved ages, metallicities, and
abundance ratios (Sect.~\ref{sec:sppars}),  constraints on the stellar
IMF (Sect.~\ref{sec:IMF}),  and stellar  and dynamical  mass estimates
(Sect.~\ref{sec:masses}).    We    discuss   the    implications    in
Sect.~\ref{sec:discussion}    and    summarize    our    results    in
Sect.~\ref{sec:summary}. Throughout this work, we adopt a $\Lambda$CDM
cosmology  with $\Omega_{\rm  m}=0.3$ and  $\rm H_0  = 70$~$\mathrm{km
  s^{-1}   Mpc^{-1}}$.    Magnitudes   are   provided   in    the   AB
system~\citep{OkeGunn:1983}.

\section{Data}
\label{sec:data}
\begin{figure}
 \begin{center}
\leavevmode
    \includegraphics[width=9cm]{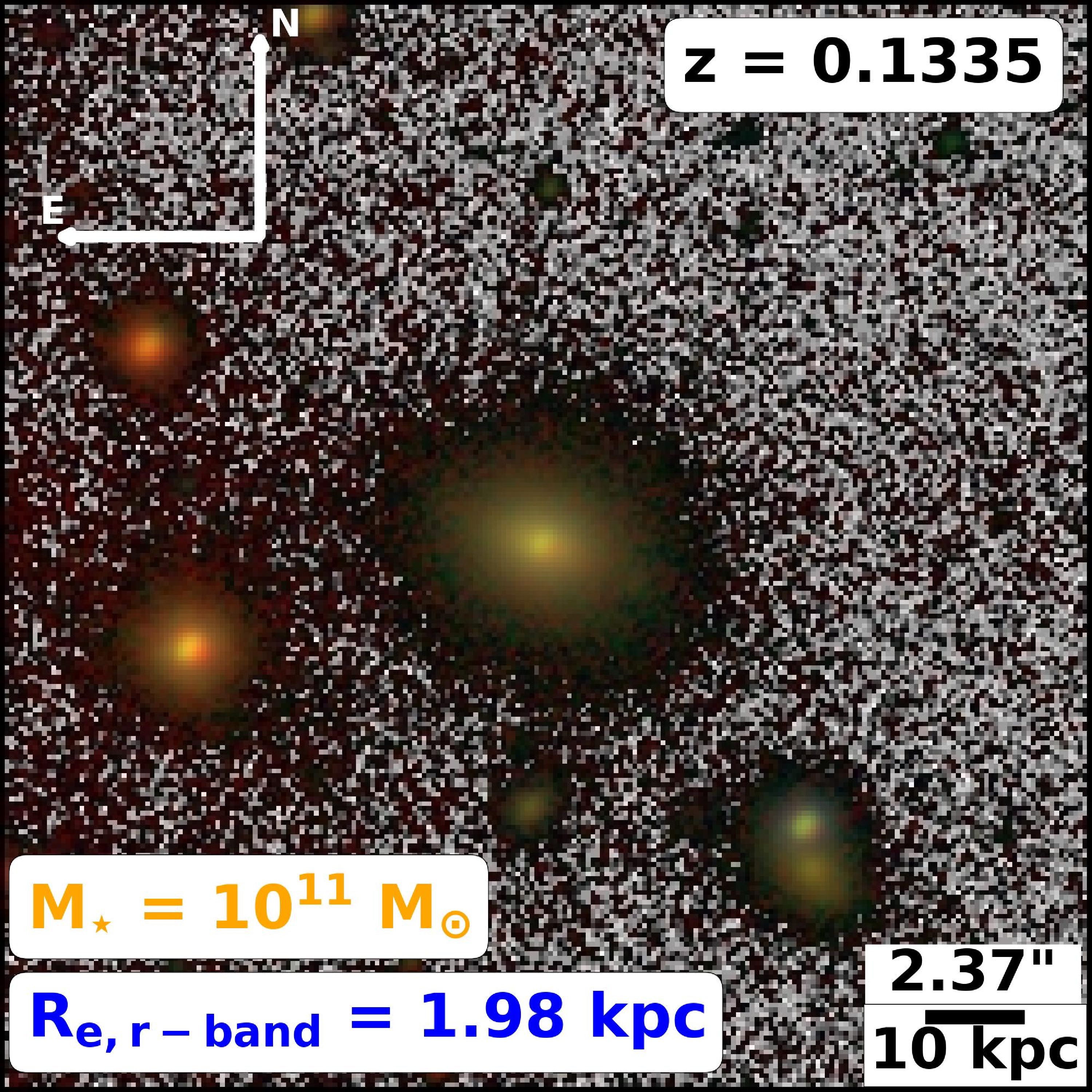}
 \end{center}
    \caption{
RGB composite image of G\,79071. The RGB channels correspond to the i-, r- and g-bands respectively from the KiDS survey \citep{deJong17}. The galaxy is extremely compact, with the characteristic swollen disk morphology from the majority of red nuggets/relic galaxies. In addition, there is a lack of nearby bright companions.
    }
    \label{fig:g79017_rgb}
\end{figure}

G\,79071  was selected  for deep  spectroscopic observations  from the
sample     of     massive     ultra-compact     galaxies     presented
in~\citet{Buitrago:2018}, owing  to its  high stellar  mass (10$^{11}$
M$_{\odot}$), small effective radius  ($R_{\rm e,r-band}$ = 1.98 $\pm$
0.11  kpc), and  relatively bright  apparent magnitude  (17.83, 16.92,
16.50,  and 16.18  AB  mag in  the $g$-,  $r$-,  $i$-, and  $z$-bands,
respectively).  The latter  reflects  the fact  that  G\,79071 is  the
lowest-redshift object  in the parent sample  ($z_{\rm spec}=0.1335$),
while its  compactness makes  this galaxy a  3$\sigma$ outlier  in the
local mass--size relation~\citep{Lange:2015}. {Taking the values published in this reference for this scaling relation, and assuming the same uncertainties as for the mass-size relation derived for the Sloan Digital Sky Survey \citep[SDSS; ][]{Shen:03}, G\,79071 is a 3.26$\sigma$ outlier for disks in the $g$-band and 4.14$\sigma$ in the $z$-band, while it is a 2.94$\sigma$ outlier for spheroids in the $g$-band and 2.86$\sigma$ in the $z$-band. Both relations are given, for disk and spheroidal morphologies, due to the well-known fact that red nuggets/relic galaxies display S0/swollen disk-type visual morphologies \citep[e.g.][]{Buitrago:2013,Buitrago:2014,Tortora:2025}}. Fig. \ref{fig:g79017_rgb}  shows a RGB
colour      image      of      G\,79071     created      with      the
\texttt{astscript-color-faint-gray}    program~\citep{Infante-Sainz24}
within the \textsc{gnuastro} library~\citep{Akhlaghi15}.

We obtained new, deep long-slit spectroscopy of G\,79071
with the X-Shooter spectrograph at the ESO-VLT, on Cerro Paranal
(Proposal ID: 105.206U, PI: IF).  X-Shooter is a
second-generation ESO-VLT instrument -- a slit echelle spectrograph
that covers a wide spectral range (3000--25000\,\AA), at relatively
high resolution~\citep{Vernet:2011}. The spectral range is covered by splitting the incoming bin into three independent arms, ultraviolet-blue (UVB: 3000--5900\,\AA);
visible (VIS: 5300--10200\,\AA); and near-infrared (NIR: 9800--25000\,\AA). For the present work, we analyze only data in the UVB and VIS arms.

The X-Shooter slit is 11'' long, with a spatial scale of 0.16\,$\rm
arcsec/pixel$ in the UVB and VIS, and 0.21\,$\rm arcsec/pixel$ in the
NIR, arms. For all observations, we adopted an instrument setup with 0.9''-, 0.9''-, and 1.0''- wide slits, resulting into a resolution power of $R \sim\!4400$, $\sim\!7500$, and $\sim\!5500$, in the UVB, VIS, and NIR arms, respectively. The average seeing of the observations was 0.7~\arcsec\ (FWHM) in the optical.

The data were reduced with a similar procedure as in \citet[][hereafter LB16 and LB17, respectively]{LB:16,LB:17}. For each arm, the data were pre-reduced using
version 2.4.0 of the data-reduction pipeline~\citep{Mod:2010},
performing the subsequent reduction steps (i.e. flux calibration, sky
subtraction, and telluric correction) with dedicated FORTRAN software
developed by the authors.

\section{Stellar population models}
\label{sec:models}
Our analysis is based on E-MILES~\citep{Vazdekis:2016} and \citet[hereafter CvD18]{CvD18} stellar population models.

E-MILES  simple stellar population (SSP) models span the spectral range from $0.168$ to 5~$\mu$m, and are based on the NGSL~\citep{Gregg:2006}, MILES~\citep{MILESI}, Indo-US~\citep{Valdes04}, CaT~\citep{CATI} and IRTF~\citep{IRTFI,IRTFII} empirical stellar libraries (see also~\citealt{Vazdekis:12},~\citealt{RV:16}). 
We use E-MILES models computed for two sets of scaled-solar theoretical isochrones: those of \citet{Padova00} (Padova00; hereafter ``iP'') and \citet{Pietrinferni04} (BaSTI; hereafter ``iT''), the latter having lower temperatures  at the low-mass end~(see \citealt{Vazdekis:15}, and references therein). The SSPs are computed for ages from $\sim 0.06$ to $\sim 17.8$\,Gyr ($0.03$ to $14$~Gyr), and metallicity, \zh, from $-2.2$ to $+0.22$~dex ($-1.7$ to $+0.26$~dex), for Padova00 (BaSTI)~\footnote{BaSTI models are also available for \zh$=0.4$ (see \citealt{Vazdekis:2016}), but these models have lower quality, and hence are not used in the present work.} isochrones. We use models computed for a ``bimodal'' initial mass function (IMF), i.e. a single power-law distribution whose logarithmic slope, $\rm \Gamma_b$, is tapered at the low-mass end ($\lesssim 0.5 \, M_\odot$). In this parametrization, increasing the high-mass end slope does also increase the dwarf-to-giant ratio in the IMF (implying a more bottom-heavy distribution) through its overall normalization. For \gammab$=1.3$, the bimodal IMF is very similar to the Kroupa IMF. 

CvD18 models are an updated version of those by~\citet[hereafter CvD12]{CvD12a}. They cover the spectral range from 0.35 to 2.5~$\mu$m and are based on the MILES and extended IRTF~\citep{Villaume:2017} empirical stellar libraries, in the optical and NIR, respectively. The models adopt MIST isochrones~\citep{Choi:2016, Dotter:2016}, and span ages from 1 to 13.5~Gyr, and metallicities from -1.5 to 0.2~dex. The SSPs adopt a three-segment IMF, with variable slopes, $\rm x_1$ between 0.1--0.5~$\rm M_\odot$, $\rm x_2$ between 0.5--1~$\rm M_\odot$, and a fixed Salpeter slope at higher masses.

CvD18 also provide a set of theoretical SSPs, computed for a Kroupa IMF, at ages of $1, 3, 5, 9$ and 13~Gyr, and metallicities of $-1.5$,  $-1$, $-0.5$, $0$, and $+0.2$~dex, where the abundance of individual elements is varied independently. In this work, we focus on elements that produce most significant variations in the optical spectral range, namely the $\alpha$-elements (O, Mg, Si, Ca, Ti), as well as the elements C, N, and Na. Notice that CvD18 models vary O, Ne, and S in lockstep, denoting the resulting abundance ratio as [$\alpha$s/Fe]. Here, we adopt the same convention, assuming that O dominates the spectral response, i.e. [O/Fe]=[$\alpha$s/Fe].
For each element X, CvD18 compute theoretical SSP spectra, $\rm S_X$, for reference abundance ratios $\rm [X/H]_r=\pm0.3$~dex, with exceptions for C (varied by $\rm \pm 0.15$~dex), and Na (varied by $\pm 0.3, +0.6$ and $+0.9$~dex, respectively). We estimate the spectral response to a given abundance ratio \xh\ through { linear interpolation:
\begin{equation}
\rm \mathcal{R}_{X} = 1 + \frac{[X/H]}{[X/H]_r} \cdot \left( \frac{S_X}{S_{\odot}} - 1 \right) ,
\label{eq:resp}
\end{equation}
where $\rm S_{\odot}$ is the scaled-solar theoretical spectrum. For a given age and metallicity, $\rm \mathcal{R}_X$ is computed by linear interpolation over the grid of CvD18 theoretical models. }

Multiplying the { responses $\rm \mathcal{R}_X$ for} the empirical SSP (either E-MILES or CvD18), we correct it to a given abundance pattern. We note that stellar libraries used in the computation of E-MILES and CvD18 SSPs follow the abundance pattern of the Milky Way (MW), i.e. they are approximately scaled-solar at solar and super-solar metallicity, and significantly $\alpha$--enhanced at metallicities below $\sim -0.2$~dex.
In practice,  all spectra analyzed in this work have metallicities below this threshold. Therefore, no correction to the inferred \xfe\ abundances is required for the MW pattern. An exception is that of O abundance, as in the solar neighborhood \ofe\ decreases with metallicity, becoming negative in the super-solar metallicity regime~\citep{Bensby:2004}, which is relevant for the innermost region of G\,79071. We discuss this issue in Sec.~\ref{sec:results}.

To extract the kinematics of G\,79071 (Sect.~\ref{sec:kin}), we also used SSP models based on the X-Shooter Spectral Library (XSL; \citealt{Verro:2022a}). The XSL SSPs cover the wavelength range $0.3$--$2.5\,\mu$m and have a higher spectral resolution ($R\sim10\,000$) than the E-MILES and CvD18 models. In the present work, we adopt XSL SSPs computed with PARSEC/COLIBRI isochrones (see \citealt{Verro:2022b} and references
therein), spanning ages from $0.05$ to $15.8$~Gyr and metallicities from $-2.2$ to $+0.2$~dex, and assuming a Kroupa IMF.

\section{Analysis}
\label{sec:methods}

\subsection{Radial binning procedure}
\label{sec:binning}
In order to extract the kinematics of G\,79071, namely the profiles of radial velocity, \vrad , and velocity dispersion, $\sigma$, we radially binned the X-Shooter two-dimensional spectra along both sides of the
slit. To account for seeing effects, we set the central bin (around the photometric center of the galaxy) to be $1.0\arcsec$ wide, while for the other bins we adapted the bin width to ensure a minimum signal S/N ratio (per \AA), $\rm S/N_{min}$, of $\sim10$. This resulted in a set of 11 binned spectra, out to a galactocentric distance of $\rm R \sim 3.25\arcsec$ (i.e. $\sim 4\,R_e$) from the centre.

For the stellar population analysis, we used the radial velocity profile to correct the two-dimensional spectra into the restframe along the slit, and extracted another set of radially binned spectra. We folded up both sides of the slit in order to achieve a higher threshold of $\rm S/N_{min} \sim 40$. We adopted this higher S/N threshold to derive the radial profiles of stellar population properties, namely \age , metallicity, \feh , and
abundance ratios, \xfe . This procedure produced a set of four binned spectra, out to a galactocentric distance of $\rm R \sim 1.75\arcsec$ (i.e. $\sim 2\,R_e$) from the centre.

Finally, to constrain the stellar IMF, we repeated the same procedure, but adopting an even higher threshold of $\rm S/N_{min} \sim 80$. This is because the effect of varying the IMF is subtle, and can be measured only with sufficiently high S/N ratios \citep[see, e.g.,][hereafter LB13]{LB:13}. We obtained two spectra with this S/N requirement, namely an ``innermost'' one ($\rm R < 1\,R_e$) and an ``outermost'' one ($\rm 1 < R < 3\,R_e$).

{ The different radial binning schemes adopted in this work are summarized in Tab.~\ref{tab:binning}.}

\begin{table*}
\caption{ Summary of the radial binning schemes adopted in this work.}
\label{tab:binning}
\centering
\begin{tabular}{lcccc}
\hline\hline
Analysis & No. spectra & Radial extent & S/N$_{\rm min}$ & Quantities measured \\
& & & per \AA & \\
\hline
Kinematics & 11 & $R \lesssim 3.25\arcsec$ ($\sim4 \, R{\rm e}$) & $\sim10$ & $V_{\rm rad}$, $\sigma$ \\
Stellar populations & 4 & $R \lesssim 1.75\arcsec$ ($\sim2 \, R_{\rm e}$) & $\sim40$ & Age, \feh, \xfe \\
IMF & 2 & $R<1 \, R_{\rm e}$ and $1<R/R_{\rm e}<3$ & $\sim 80$ & IMF slope, mass-excess factor \\
\hline
\end{tabular}
\tablefoot{ The radial extents in kpc are computed using the angular
scale at the redshift of G\,79071, namely $2.37\,{\rm kpc\,arcsec^{-1}}$. For the stellar population and IMF analyses, the two sides of the slit are folded together to increase
the S/N.}
\end{table*}

\subsection{Extraction of spectroscopic properties}
\label{sec:properties}

\subsubsection{Kinematics}
We measured the kinematics of G\,79071, namely the radial velocity, \vrad, and velocity dispersion, $\sigma$, using the {\sc pPXF} software \citep{Cap:2004, Capp:17}. The pPXF fits were performed over the wavelength range $3900$--$5400$~\AA, which includes the most prominent optical absorption features (e.g. Mgb, Fe lines, Balmer lines, and the G band). We ran pPXF including an additive polynomial of degree 10. Uncertainties on the kinematics were estimated by varying the polynomial degree by $\pm3$, as well as by shifting each end of the fitted wavelength range by $\sim10\%$. We note that, for E-MILES, the model resolution of $2.5$~\AA\ (FWHM) corresponds to an instrumental dispersion of $\sigma_{\rm inst}\sim80$~\kms\ ($\sim60$~\kms) at the blue (red) end of the fitted range. Therefore, before running pPXF we convolved all the E-MILES SSPs to a constant dispersion of $\sigma_{\rm const}\sim80$~\kms, and then, for each spectrum of G\,79071, we obtained the velocity dispersion by subtracting $\sigma_{\rm const}$ in quadrature from the dispersion measured by pPXF. We repeated the same procedure when running pPXF with the XSL models. However, in this case the model resolution ($\sim15$~\kms\ FWHM in the optical) is negligible compared to the galaxy velocity dispersions at all radii, making any correction for $\sigma_{\rm const}$ unnecessary. For this reason, in the outermost bins of G\,79071, where $\sigma\sim60$--$80$~\kms\ (i.e. comparable to $\sigma_{\rm inst}$ for the E-MILES models over the fitted wavelength range), we relied entirely on the XSL-based estimates, as shown below (Sect.~\ref{sec:kin}).

\subsubsection{Stellar population properties}

Stellar  population  properties,  \age   ,  metallicity,  \feh  ,  and
abundance ratios, \xfe  , were derived following the  same approach as
in   \citet[][hereafter  LB26]{LB:26},   by  combining   results  from
different fitting  techniques and by  considering two sets  of models,
namely  E-MILES  SSPs with  either  Padova  or BaSTI  isochrones  (see
Sect.~\ref{sec:models}) and the CvD18 models. In practice, we computed
a  final  estimate  at  each  radius by  averaging  the  results  from
complementary  approaches:  full  spectral fitting  (FSF),  full-index
fitting (FIF), and index fitting (IF):
\begin{description}
\item[FSF  --] The  best-fitting  model spectrum  is obtained  through
  $\chi^2$ minimization of the flux residuals between data and models,
  while accounting for  continuum-shape mismatches with multiplicative
  polynomials. Following  the same approach as  in~\citet{CvD12b}, the
  fit  is performed  by  splitting the  spectra  into four  wavelength
  intervals,   { intervals, $\Delta(\lambda)=$4000--4900~\AA, 4950--5800~\AA, 5800--6400~\AA, and 8000--8750~\AA.}
\item[FIF --] The same $\chi^2$ minimization in flux space is carried out, but the
fit is restricted to the passbands of a set of Lick-like { indices,
after normalizing both the observed and model spectra by their
pseudo-continua~\citep{NMN:19} and excluding indices affected by the X-Shooter
dichroic, data-reduction artifacts, or sky residuals (see below).}
\item[IF --] Parameter constraints are obtained by fitting the equivalent widths of the same indices, minimizing the difference between observed and model line strengths, including the associated index uncertainties.
\end{description}

{ For the FIF and IF methods, we use the following set of spectral
indices: the age-sensitive Balmer indices \hgf\ and
\hdf;\footnote{We did not include \hb\ in the index-fitting procedure
because the red pseudo-continuum of this feature partly overlaps with
the X-Shooter dichroic region, making it less robust than the other
indices. We verified, however, that including \hb\ does not
significantly affect our results. The \hb\ region is included in the
FSF approach by avoiding the region most affected by the dichroic; see, for example, App.~\ref{app:fits}.} the iron indices
Fe4383, Fe4531, Fe5015, Fe5270, and Fe5335; the sodium indices \nai\
and \naii; the magnesium indices \mgf\ and \mgb; the calcium indices
\cafr, \cai, and \caii; and additional indices including \tioi,
\tioiir, C4668, \cnii, G4300, Si4101, and Si4513. For most indices,
the central passband and pseudo-continuum definitions follow the Lick
system \citep{Trager98}. The exceptions are \cafr, for which we adopt
the improved Ca4227 definition of \citet{PRS:05}; \tioiir, based on
the modified \tioii\ definition of~LB13; \mgf, Si4101, and
Si4513, defined as in \citet{Serven:2005}; and \naii, defined as in
\citet{CvD12a}, with the modifications described in~LB13.}

We incorporated the effect of elemental abundance variations, \xfe , with $X=\{{\rm O,\,Mg,\,Si,\,Ca,\,Ti,\,C,\,N,\,Na}\}$, by applying response functions to each SSP spectrum (see Sect.~\ref{sec:models}, and LB26 for details). For all methods, we adopted models consisting of a linear combination of two SSPs, with different age and metallicity, but sharing the same IMF and abundance mixture (i.e. the same set of \xfe ). In practice, for G\,79071 we find only negligible changes in the inferred parameters when adopting a single SSP.
Prior to fitting, all models were convolved to match the velocity dispersion, $\sigma$, of each observed spectrum.
No contamination from nebular emission was detected in the spectra, and hence no emission correction was applied.
Uncertainties on the best-fitting parameters were estimated via Monte Carlo realizations of the input spectrum, by
perturbing the fluxes according to their formal errors. The quality of the fits for G\,79071 turned our to be excellent for all spectra (with rms residuals at the sub-percent level for both FSF and FIF), as illustrated in
App.~\ref{app:fits}. The final estimates of \age , \feh , and \xfe\ were obtained by combining the results from all three fitting methods (FSF, FIF, and IF). For each method, the fits were repeated for different IMF shapes, and the results were then marginalized over the IMF. We note, however, that the inferred stellar population parameters are largely unchanged when the IMF is fixed (e.g. to a Kroupa-like distribution).

IMF slopes were constrained using only the FIF and IF approaches, in order to maximize the leverage of the
IMF-sensitive features included in our analysis (i.e. \mgf, \nad, the TiO bands, \naii, and the calcium triplet lines, Ca1 and Ca2; see LB26 for details). When fitting the spectra with the E-MILES models, we adopted the low-mass tapered (``bimodal'') IMF, parametrized by the slope \gammab\ (see Sect.~\ref{sec:models}). For CvD models, the low-mass IMF slopes ($\rm x_1$ and $\rm x_2$; see Sect.~\ref{sec:models}) were treated as free parameters. Finally, for each fit, the IMF constraints from each set of models were converted into a mass-excess factor, $\alpha$, defined as the $r$-band luminosity-weighted best-fitting \mlr\ normalized to the value expected for a fixed, Kroupa-like IMF.

\section{Results}
\label{sec:results}

\subsection{Kinematics}
\label{sec:kin}

\begin{figure}
 \begin{center}
\leavevmode
    \includegraphics[width=9cm]{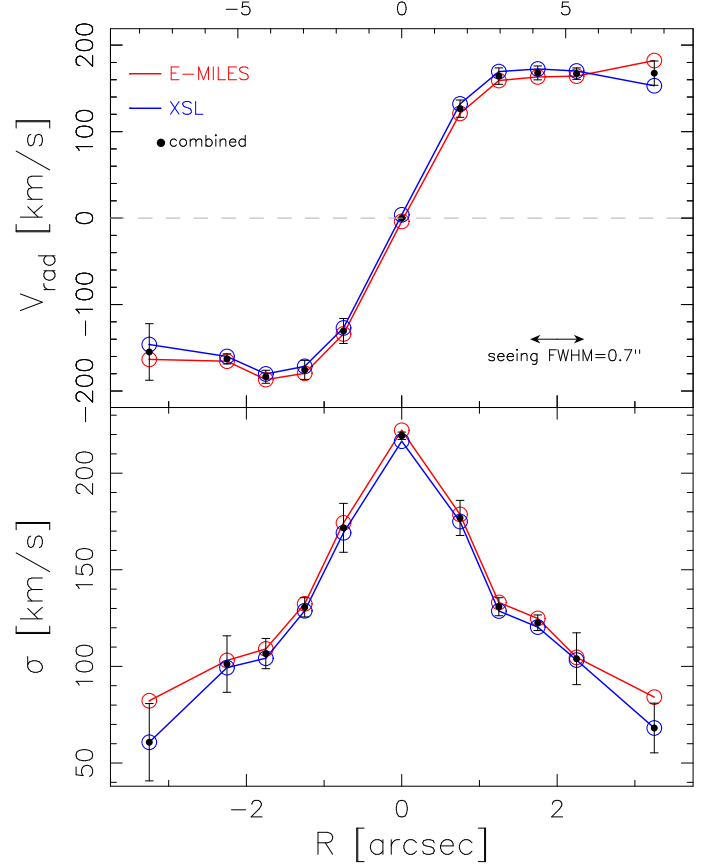}
 \end{center}
    \caption{
Kinematics of G\,79071. Upper and lower panels show the radial velocity, $\rm V_{rad}$, and velocity dispersion, $\sigma$, measured along the X-Shooter slit as a function of galactocentric distance, $\rm R$. Red and blue curves correspond to the pPXF kinematics obtained using the E-MILES and XSL stellar population models, respectively. Black points with error bars show the combined measurements adopted in this work (see text). The average seeing FWHM of the X-Shooter observations was $0.7$~\arcsec , as shown in the lower-right corner of the upper panel.
}
    \label{fig:kin}
\end{figure}

Fig.~\ref{fig:kin} shows that G\,79071 exhibits significant rotation, with \vrad\ increasing up to $\sim 180$~\kms\ at galactocentric distances larger than $\sim 2$~kpc ($>1\,\rm R_e$), where the rotation curve reaches a plateau. The velocity dispersion, $\sigma$, decreases significantly from a central value of $\sim 220$~\kms\ down to $\sim 70$~\kms\ at the outermost radius probed in this work ($\rm R \sim 8$~kpc).

The central value of $\sigma$ is fully consistent with the SDSS fibre-integrated measurement, $213 \pm 8$~\kms. We note that, at all radii, the kinematics derived from the E-MILES and XSL models are fully consistent, with the largest discrepancy occurring in the outermost bin, where the XSL models yield a lower $\sigma$ by $\sim 20$~\kms. This difference is likely due to the fact that, in the outermost bin, the intrinsic resolution of the E-MILES models is $\sim 80$~\kms\ (see above), i.e. comparable to the galaxy velocity dispersion. For this reason, we derived the final kinematics (black dots in Fig.~\ref{fig:kin}) by averaging the results obtained with the XSL and E-MILES models, except for $\sigma$ in the outermost bin, for which we adopted the XSL value only.

Assuming that the kinematic profiles share the same symmetry { as the KiDS photometry of G\,79071, with an axis ratio of $b/a\sim0.7$ \citep{Buitrago:2018},  we estimate $(V/\sigma)_e \sim 0.46$. Given the corresponding ellipticity, $\epsilon = 1-b/a \sim 0.3$, this value indicates substantial rotational support and is consistent with a fast-rotator-like kinematic structure \citep{Cappellari07}.  However, this interpretation is based on long-slit data and not on a two-dimensional kinematic classification.}

\subsection{Age, metallicity, and abundance ratios}
\label{sec:sppars}

\begin{figure}
 \begin{center}
\leavevmode
    \includegraphics[width=8cm]{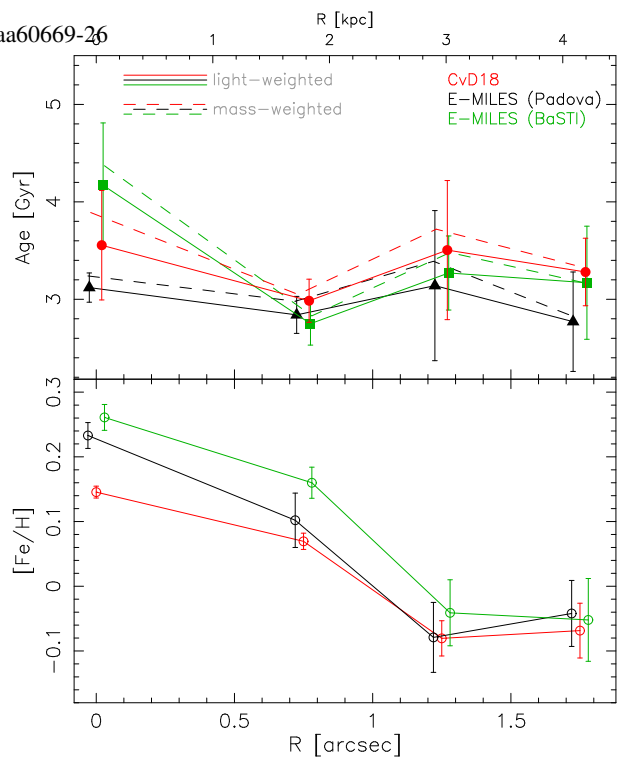}
 \end{center}
    \caption{
    Age (top) and metallicity, \feh,  profiles of G\,79071 as a function of galactocentric distance, $\rm R$. Red, black, and green curves show results obtained with the CvD18 models and the E-MILES models based on Padova and BaSTI isochrones, respectively. Solid and dashed curves correspond to luminosity- and mass-weighted estimates. Error bars denote $1\sigma$ uncertainties.
    }
    \label{fig:agez}
\end{figure}

Figure~\ref{fig:agez} shows that, regardless of the adopted set of
stellar population models, G\,79071 exhibits a nearly flat age
profile, with $\age \sim 3$~Gyr at all radii probed by our data. {
  As shown in App.~\ref{app:age_z_deg}, the age-metallicity likelihood contours
  obtained from full spectral fitting are well localized and do not
  extend toward significantly older ages, indicating that the inferred
  intermediate age is not driven by the age-metallicity degeneracy.}
We also find only a small difference ($\lesssim 0.2$~Gyr) between
luminosity- and mass-weighted ages, implying that the galaxy does not
host a significant old stellar component.  To further test this point,
we repeated the fit by including a third SSP component, forcing its
age to be older than 5~Gyr. { 
For all adopted stellar population models, the best-fitting contribution of
this old component remained below 10 per cent (see App.~\ref{app:age_z_deg}).}
The metallicity profile of G\,79071 (lower panel of
Fig.~\ref{fig:agez}) displays a clear negative radial gradient.  A
linear fit of \feh\ as a function of $\rm \log R$ yields a logarithmic
gradient of $\sim-0.2$~dex per radial decade, consistent with typical
values reported for ETGs \citep[e.g.,][]{LB:12,IF:19}.

\begin{figure}
 \begin{center}
\leavevmode
    \includegraphics[width=8cm]{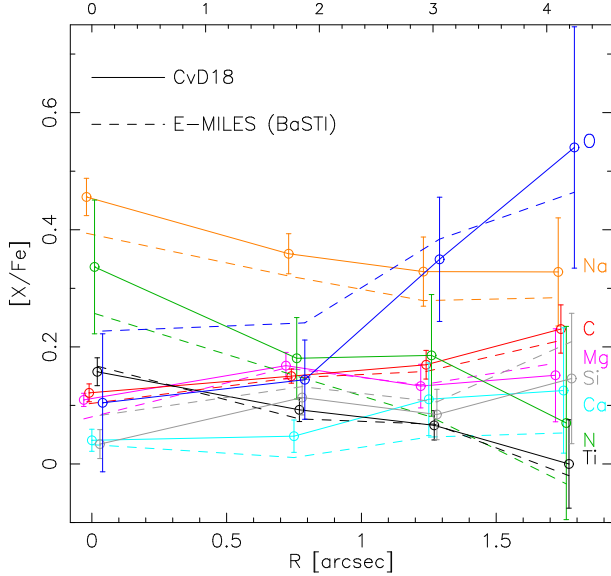}
 \end{center}
    \caption{
    Individual abundance ratios, \xfe , for G\,79071 as a function of galactocentric distance, $\rm R$. Different colours indicate different elements (labels on the right). At a given radius, small horizontal offsets are applied between elements for clarity. Solid and dashed curves show results obtained with CvD18 models and E-MILES models based on BaSTI isochrones, respectively. Error bars denote $1\sigma$ uncertainties.
    }
    \label{fig:AB}
\end{figure}

Figure~\ref{fig:AB} shows the radial profiles of the individual abundance ratios, \xfe , for G\,79071, comparing the results obtained with the CvD18 models (solid lines) to those from the E-MILES models based on BaSTI isochrones (dashed lines). Overall, the abundance profiles are fairly flat and broadly consistent between the two model sets~\footnote{Somehow this might be expected, as in all cases we model  the effect of abundance ratios using the same set of response functions (from CvD18).}. The $\alpha$ elements (e.g. Mg and O) are mildly enhanced in the central regions, at the level of $\sim 0.1$~dex, while Na and N exhibit the largest enhancements, reaching up to $\sim0.4$~dex. We note that \ofe\ appears to increase with galactocentric distance; however, oxygen is notoriously difficult to constrain, as also reflected by the relatively large uncertainties on \ofe , especially in the outermost radial bins. 
{ Moreover, as discussed in Sec.~\ref{sec:methods}, abundance ratios in our approach are measured relative to the abundance pattern of MW stars. This is particularly relevant for \ofe, since MW stars show decreasing \ofe\ at super-solar metallicity~\citep{Bensby:2014}. Given the uncertainties affecting the MW trend itself, we do not correct our measurements, but estimate the possible impact on our results. We approximate the trend of MW disk stars in Fig.~15 of \citet{Bensby:2014} with the linear relation $\rm [O/Fe]_{MW}=-0.38\,[Fe/H]_{MW}$. For the two innermost spectra of G\,79071, which have super-solar \feh, applying the corresponding MW correction would decrease the inferred \ofe\ by $\sim0.08$ dex and $\sim0.05$ dex, respectively. This would not affect our main interpretation, strengthening the conclusion that \ofe\ increases with radius in G\,79071.}

\begin{figure}
 \begin{center}
\leavevmode
    \includegraphics[width=8cm]{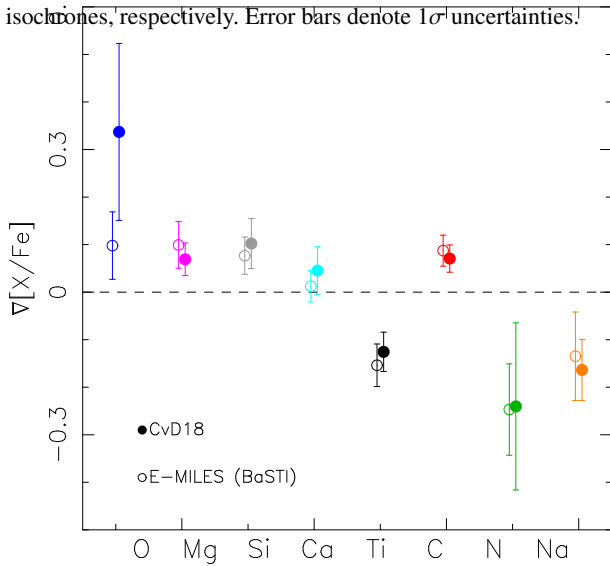}
 \end{center}
    \caption{
    Logarithmic radial gradients of individual abundance ratios, $\nabla[X/{\rm Fe}]$, for different elements (see horizontal axis), in G\,79071. Filled and open symbols refer to results for CvD18 models and E-MILES models based on BaSTI isochrones, respectively. Error bars denote $1$-sigma uncertainties. The horizontal dashed line marks a value of zero.
    Different colours indicate different elements, as in Fig.~\ref{fig:AB}.
    }
    \label{fig:AB_GRAD}
\end{figure}

To quantify radial trends, we fitted \xfe\ as a function of $\log R$ with a linear relation, weighting each point by its uncertainty. The resulting slopes, that is, the logarithmic radial gradients $\nabla[X/{\rm Fe}]$, are shown in Fig.~\ref{fig:AB_GRAD}. The figure indicates that the $\alpha$ elements in G\,79071 exhibit a positive radial gradient~\footnote{The only exception is Ti, which shows a negative radial gradient (see the black points in Fig.~\ref{fig:AB}). However, as shown in LB26a, the inferred \tife\ is somewhat model dependent and should therefore be interpreted with caution.}, in qualitative agreement with findings for massive ETGs (e.g. \citealt{vanDokkum:2017}). This behaviour can be interpreted as evidence for a slightly more extended star-formation timescale in the central regions compared to the outskirts, leading to comparatively lower $\alpha$ enhancement at small radii. We also find a positive gradient for C. This is consistent with previous results suggesting that C receives a substantial contribution from massive stars (as the alpha elements), as found for the bulge of M31 in LB26 and for Milky-Way stars~\citep{Romano:2020}. We note that both N and Na  show negative radial gradients in G\,79071. For Na, this result might be explained by the fact that Na is mostly produced in massive stars, with metallicity-dependent yields (see~\citealt{LB:17}).

\begin{figure}
 \begin{center}
\leavevmode
    \includegraphics[width=8cm]{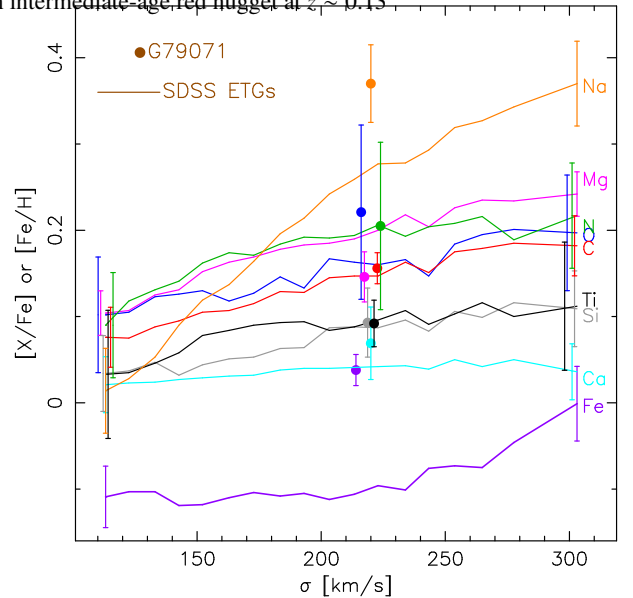}
 \end{center}
    \caption{
Abundance ratios, \xfe , and metallicity, \feh, for G\,79071 (filled circles with error bars) are compared
to those derived from stacked SDSS spectra of ETGs as a function of velocity dispersion, $\sigma$ (solid lines).
Abundances for G\,79071 are estimated by mimicking the SDSS fiber aperture (see text).
    }
    \label{fig:SDSS_AB}
\end{figure}

Figure~\ref{fig:SDSS_AB} compares the abundance ratios and metallicity
of G\,79071 with those inferred from stacked spectra of SDSS ETGs in
LB26. The SDSS stacks were analysed using the same fitting procedure
and the same SSP models adopted for G\,79071. 
{ Since spatially
resolved spectroscopy is not available for the SDSS stacks, no aperture
correction was applied to them. Instead, to enable a consistent
comparison, we report for G\,79071}  the values of \xfe\ and
\feh\ measured by mimicking the SDSS fibre aperture, that is, by
averaging the radial measurements along the X-Shooter within the SDSS
fibre radius ($1.5$~\arcsec ). The average is computed by weighting
each radial point by the flux enclosed in the corresponding elliptical
annulus at that radius. The figure shows that the chemical signatures
of G\,79071 are broadly similar to those of ``normal'' (non-compact)
ETGs at comparable velocity dispersion ($\sim 220$~\kms ).  The main
differences are in \feh\ and \nafe , which are significantly higher
(at the $\gtrsim 2\sigma$ level) than in the average SDSS population.
This points to an efficient chemical enrichment in G\,79071, with a
rapid build-up of the iron content, while the Na enhancement is likely
driven primarily by metallicity-dependent stellar yields.

{ We note that the SDSS stacks mostly consist of normal-sized ETGs.
Therefore, the differences between G\,79071 and these stacks may
reflect either the compact nature of G\,79071, a different formation
history, or a combination of both effects. A direct assessment of this
issue would require high-quality, high-S/N spectra for a representative
sample of compact ETGs with similar velocity dispersion and effective
radius, which is beyond the scope of the present paper.}

\subsection{Stellar IMF}
\label{sec:IMF}

\begin{figure}
 \begin{center}
\leavevmode
    \includegraphics[width=8cm]{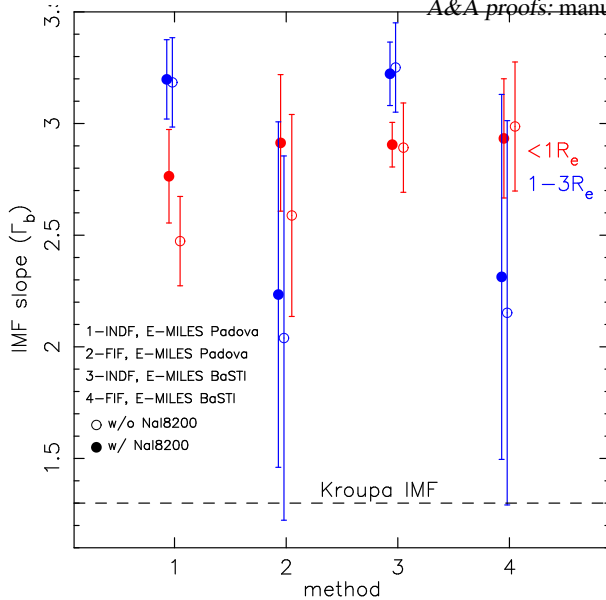}
 \end{center}
    \caption{
    Low-mass tapered (bimodal) IMF slope inferred for G\,79071 using different methods (see legend and labels on the horizontal axis). Red and blue symbols refer to the innermost ($\rm R<1\,R_e$) and { outermost ($\rm 1<R<3\,R_e$) apertures}, respectively. Error bars denote $1$-sigma confidence intervals. Filled and open symbols are obtained when including and excluding the NaI8200 spectral feature, respectively. The dashed horizontal axis corresponds to the case of a Kroupa-like IMF. 
    }
    \label{fig:IMF}
\end{figure}

Figure~\ref{fig:IMF} shows the best-fitting IMF slope obtained with the E-MILES SSP models, adopting the low-mass tapered (``bimodal'') IMF parametrized by \gammab\ (see Sect.~\ref{sec:models}). We report results derived with the FIF and INDF fitting methods and for models based on different isochrones (see the labels on the $x$-axis and the figure legend). We further test the robustness of our IMF constraints by repeating the analysis
excluding the \naii\ absorption feature (open symbols in Fig.~\ref{fig:IMF}). Since \naii\ is among the most sensitive indicators of very low-mass stars in our spectral range, it can have a significant impact on the inferred IMF slope and mass-to-light ratio (see, e.g., \citealt{CvD12b}).

In the innermost radial bin, within one effective radius, we find that the IMF of G\,79071 is significantly bottom-heavy, with \gammab$\simeq 2.5$--3, comparable to values inferred for the most massive ETGs (with $\sigma \gtrsim 260$~\kms; see, e.g., LB13; \citealt{Spiniello:2014}). This conclusion is independent of the fitting method and of the adopted set of stellar population models. Remarkably, excluding \naii\ yields IMF slopes fully consistent with those obtained when all features are included, supporting the robustness of our result.

In the outermost radial bin, between two and three effective radii,
the line-strength fitting (INDF; see labels 1 and 3 in
Fig.~\ref{fig:IMF}) points to an IMF similarly bottom-heavy as in the
center, with \gammab$\gtrsim 3$. In contrast, the FIF approach yields
substantially larger uncertainties, with IMF slopes consistent with
both a Kroupa-like and a bottom-heavy distribution.  { This is
  likely because full-spectral fitting methods, including FIF, combine
  the contribution of all spectral pixels, many of which have little
  direct sensitivity to the IMF. As a result, the IMF signal can be
  diluted, especially in outer radial bins where the spectral
  constraints are weaker.  On the other hand, } line-strength fitting
maximizes the leverage of the IMF-sensitive indicators in the $\chi^2$
computation. { In particular, in the outermost bin the measured
  \naii\ equivalent width is $\sim0.8\pm0.1$~\AA. After accounting for
  the relevant stellar population parameters, including Na abundance
  variations, and for the appropriate broadening conditions, such a
  strong \naii\ absorption cannot be reproduced with a Kroupa-like
  IMF,~\footnote{For an SSP with solar metallicity, \Age$=3$~Gyr, a
  Kroupa-like IMF, and $\sigma\sim140$~\kms\ (i.e.  the velocity
  dispersion of the outermost-bin spectrum), E-MILES models predict
\naii$\simeq0.45$~\AA .  For \nafe$\sim0.35$~dex and
  \afe$\sim0.1$~dex (Fig.~\ref{fig:AB}), this implies \naii
  $\simeq0.55$~\AA, more than $2\sigma$ below the observed value.}
  and instead favours a bottom-heavy IMF (see also \citealt{F:13}). We
  note that \naii\ is also sensitive to Na abundance, although to a
  lesser extent than to the IMF.  This effect is controlled in our
  fitting procedure by fitting \naii\ together with NaD, the latter
  being more sensitive to Na abundance than to the IMF.}

Overall, these results { suggest} that the IMF of G\,79071 is
bottom-heavy at all radii. In this respect, the galaxy is similar to
NGC\,1277, the prototype compact massive ETG for which a detailed
radial IMF analysis has been presented by~\citet{NMN:15c}. As shown in
the following section, we obtain consistent conclusions when fitting
the spectra with the CvD18 models, which adopt a different IMF
parametrization (a two-segment, rather than a low-mass tapered form).

\begin{figure*}
\sidecaption
  \includegraphics[width=11cm]{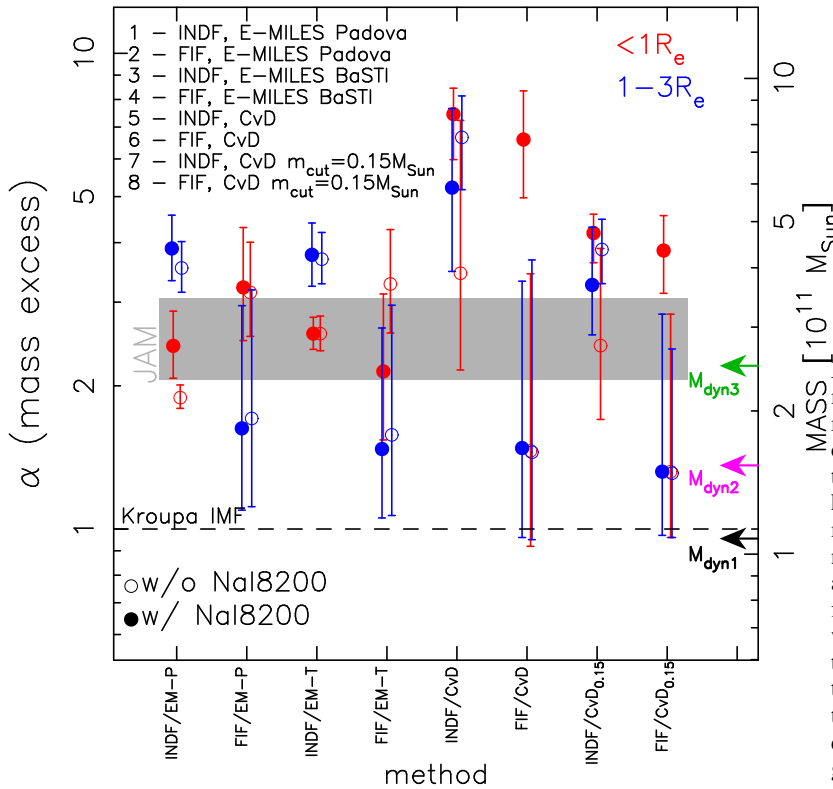}
    \caption{
    Mass-excess factor, $\alpha$, (left axis) and stellar mass (right axis) inferred for G\,79071 using different methods and stellar population models. { The numerical labels 1--8 on the upper horizontal axis correspond to the methods listed in the legend, while the lower horizontal axis reports the corresponding abbreviated method names}. Red and blue symbols refer to the innermost ($\rm R<1,R_e$) and outermost ($\rm 1<R<3,R_e$) apertures, respectively. Error bars denote $1$-sigma confidence intervals. Filled and open symbols are obtained when including and excluding the NaI8200 spectral feature, respectively. The horizontal dashed line indicates the expectation for a Kroupa-like IMF. Small horizontal arrows on the right-hand side mark dynamical mass estimates from different virial estimators (see text). The grey shaded region shows the range obtained by fitting the kinematics with JAM models assuming a range of plausible inclinations (see text).
    }
    \label{fig:alpha_mass}
\end{figure*}

\subsection{Stellar and dynamical masses}
\label{sec:masses}

Figure~\ref{fig:alpha_mass} summarizes the stellar mass-to-light ratio constraints from our IMF analysis in terms of the mass-excess factor, $\alpha$, for different fitting techniques and stellar population models (see legend and labels on the horizontal axis), and for the innermost and outermost apertures (red and blue symbols), respectively. For CvD18 models, we also consider constraints obtained by varying the low-mass end cutoff, $m_{\rm cut}$, of the IMF, as in~\citet{Barnabe:2013} (labels 8--9 in the Figure; see below). For each fit, $\alpha$ is defined by normalizing the best-fitting $r$-band mass-to-light ratio to that expected for a Kroupa-like IMF,\footnote{Given the age and metallicity of G\,79071, we have $(M_\star/L_r)_{\rm Kr}\sim1.6$.} computed for the \emph{same} best-fitting age and metallicity:
\begin{equation}
\alpha \equiv \frac{(M_\star/L_r)_{\rm fit}}{(M_\star/L_r)_{\rm Kr}} ,
\end{equation}
so that $\alpha=1$ corresponds to a Kroupa-like IMF and $\alpha>1$ indicates an IMF heavier than Kroupa. Filled and open symbols refer to fits including and excluding the \naii\ feature, respectively. The right-hand axis of Fig.~\ref{fig:alpha_mass} shows the corresponding stellar mass, obtained by converting $(M_\star/L_r)_{\rm fit}$
into $M_\star$ using the total $r$-band luminosity of the galaxy (see Sect.~\ref{sec:data}).

The inferred $\alpha$ values closely reflect the IMF trends discussed in Sect.~\ref{sec:IMF}. In the inner aperture ($\rm R<1\,R_e$), all methods and model sets yield $\alpha>1$, implying a significantly bottom-heavy IMF. This conclusion holds for E-MILES (independent of the adopted isochrones) and is recovered as well with the CvD18 models
(labels 6--9), despite their different IMF parametrisation. In the outermost aperture, the results are more method dependent; nevertheless, the index-fitting constraints (INDF) consistently favour $\alpha>1$, { suggesting that the IMF remains bottom-heavy} also at large radii, in agreement with Sect.~\ref{sec:IMF}.

We also compare the IMF-based stellar masses with dynamical mass estimates, shown by the small horizontal arrows on the right-hand side of Fig.~\ref{fig:alpha_mass}. These comparisons are meaningful only when
$M_{\rm dyn}\gtrsim M_\star$, since $M_{\rm dyn}$ represents the total mass enclosed within the adopted aperture. We first consider virial scalings of the form $M_{\rm dyn}\propto K\,\sigma_e^2 R_e/G$ (black leftwards arrow), where the ``classical'' choice $K=5$ is widely used for local ETGs \citep{Cappellari:2006}. For G\,79071, this
estimator yields $M_{\rm dyn}<M_\star$ even for a Kroupa-like IMF (i.e. $\alpha=1$), and is therefore physically inconsistent with our stellar population analysis. This behaviour is consistent with the results of
\citet{PeraltaDeArriba:2014}, who argued that, for compact systems, the assumption of homology embedded in a fixed virial coefficient breaks down. Estimators that account for non-homology, for instance via a S{\'e}rsic-dependent virial coefficient $\beta(n)$ \citep{Bertin:2002} (pink leftwards arrow), and especially the compactness-corrected prescription proposed by \citet[their Eq.~12]{PeraltaDeArriba:2014} (green leftwards arrow), yield larger dynamical masses and provide substantially improved consistency with the stellar masses implied by a bottom-heavy IMF.

\begin{figure}
 \begin{center}
\leavevmode
    \includegraphics[width=8cm]{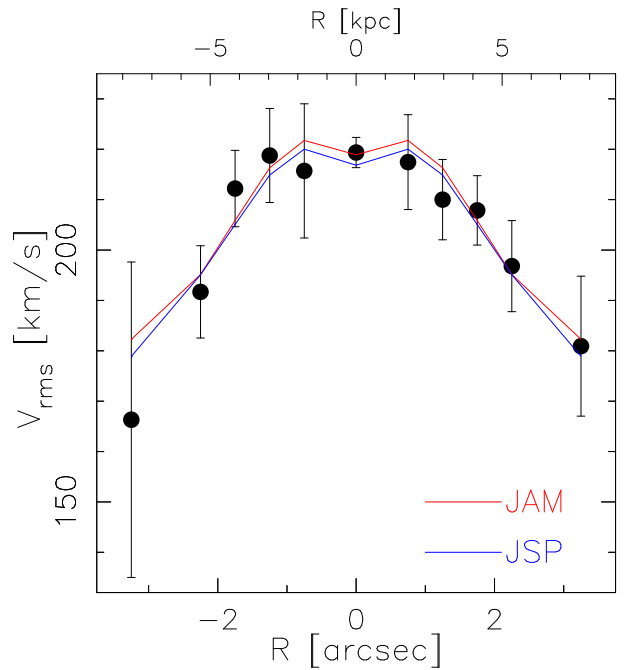}
 \end{center}
    \caption{
    Second moment of the velocity distribution, $\rm V_{rms}$, measured along the X-Shooter slit as a function of galactocentric distance, $\rm R$. Blue and red curves show the best-fitting Jeans spherical (JSP) and Jeans anisotropic (JAM) models, respectively; the JAM fit assumes an intrinsic inclination of $i=60^\circ$ (see text).
    }
    \label{fig:vrms}
\end{figure}

Finally, we compare $\alpha$ and $M_\star$ to dynamical constraints from Jeans modelling of the observed $V_{\rm rms}\equiv\sqrt{V^2+\sigma^2}$ profile (Fig.~\ref{fig:vrms}). In the spherical Jeans implementation, we
assume an NFW halo \citep{Navarro:1996,Navarro:1997} plus a spherical stellar component following a S{\'e}rsic law (see Sect.~\ref{sec:data}). For the NFW halo, we adopt the mass--concentration relation of
\citet{DuttonMaccio:2014}. The luminosity-weighted second moment of the velocity distribution is convolved with the average PSF of the X-Shooter observations and integrated along the slit within the same apertures where $V_{\rm rms}$ is measured. 

The $V_{\rm rms}$ profile is reproduced well (blue curve in Fig.~\ref{fig:vrms}); however, this requires a massive dark-matter halo, with $M_{200}\sim10^{13},M_\odot$, and a strongly negative anisotropy parameter, $\beta\sim-1.5$, which is unphysical. { This result should therefore not be interpreted as a physical constraint on the orbital structure, but rather as an indication that the assumptions of spherical symmetry and constant anisotropy are inadequate for this flattened, rotating system.} Despite this, the dynamical normalization inferred from the spherical model corresponds to $\alpha\simeq2.2$, remarkably consistent with the stellar-population inference for a low-mass tapered IMF (see labels 1--4 in Fig.~\ref{fig:alpha_mass}).

Because of the unphysical $\beta$ constraint, and { since G\,79071 is a flattened ($b/a \! \sim \! 0.7$; \citealt{Buitrago:2018}) and rotating (Sect.~\ref{sec:kin}) system}, we also performed Jeans anisotropic modelling (JAM; \citealt{Cappellari:2008}), which solves the axisymmetric Jeans equations allowing for orbital anisotropy, $\beta_z$. As for the spherical modelling, we included an NFW halo component \citep{Navarro:1996,Navarro:1997} and fit the observed kinematics along
the slit. Since we only have long-slit data, the inclination was constrained using the observed photometric
{ flattening, $Q\equiv b/a=0.7$}. Assuming axisymmetry and adopting the intrinsic flattening distribution inferred
for ATLAS$^{\rm 3D}$ fast rotators \citep{Weijmans:2014}, this corresponds to a most likely { inclination of
$i\simeq 50^\circ$.} For this value, the best-fitting JAM model reproduces the $V_{\rm rms}$ profile very well (red
curve in Fig.~\ref{fig:vrms}), { with modest anisotropy, $\beta_z=0.08\pm0.1$, a best-fitting
$M_{200}=4^{+10}_{-3.8}\times10^{13}\,M_\odot$, and $\alpha=2.3\pm0.1$.} We also explored the effect of inclination by
repeating the fit over the full allowed range of $i$ consistent with $Q$ under the same assumptions.\footnote{In
practice, the allowed range is defined by the inclinations for which the deprojection of the adopted MGE model
remains physical, given the intrinsic flattening distribution of \citet{Weijmans:2014}.} The corresponding range of
best-fitting $\alpha$ values (horizontal grey shaded region in Fig.~\ref{fig:alpha_mass}) spans { from $\sim2$ up to
$\sim 3$ for } an edge-on configuration ($i=90^\circ$).

We note that while the JAM estimates of $\alpha$ are in excellent agreement with our stellar population analysis
for a low-mass tapered IMF, the CvD18 two-segment parametrisation (labels 6--7 in Fig.~\ref{fig:alpha_mass}) tends
to yield significantly higher $M_\star/L_r$, with $\alpha>5$. The tension is partly reduced by increasing the
low-mass cutoff to $m_{\rm cut}=0.15\,M_\odot$, since stars below this mass do not significantly affect the optical
IMF-sensitive absorption features (including \naii ) while contributing non-negligibly to the mass budget. For
$m_{\rm cut}=0.15\,M_\odot$, the CvD18-based fits give $\alpha\sim4$, leaving a residual discrepancy at the
$\sim 2\sigma$ level with respect to the JAM results.

For the JAM models, the projected dark-matter fraction within one { effective radius is
$f_{\rm DM}(<R_e)=0.18\pm0.07$, where} the uncertainty includes the propagation of the inclination range allowed by
the fit. This value is consistent with dynamical studies of local ETGs, which typically find modest central dark
matter fractions within $R_e$ (e.g. a median $f_{\rm DM}\sim0.13$ within $R_e$ in ATLAS$^{\rm 3D}$;
\citealt{Cappellari:2013a}; see also \citealt{Tortora:2012}).

In summary, the JAM constraints provide a coherent picture in which the dynamical normalisation is compatible with a bottom-heavy IMF (Fig.~\ref{fig:alpha_mass}), and are in very good agreement with the non-homologous dynamical mass estimator of \citet{PeraltaDeArriba:2014}. In this respect, G,79071 differs from the prototype massive compact galaxy NGC,1277, for which dynamical studies have argued for a negligible dark-matter contribution within the stellar body \citep[e.g.][]{Yildirim:15,Comeron:2023}. { This comparison should also be interpreted in light of the markedly different environments of the two systems, as discussed in Appendix~\ref{app:ngc1277}.}

\section{Discussion}
\label{sec:discussion}

Simulations of galaxy formation within dark-matter haloes predict that compact massive galaxies that survive as relics to $\rm z\sim0$ should be found preferentially in high-density, high-mass environments (i.e. clusters; see~\citealt{QuilisTrujillo:2013, Stringer:2015}). High-velocity-dispersion environments are naturally hostile to mergers, and the hot intra-cluster medium suppresses late cold-gas accretion, providing favorable conditions for preserving compact systems over long timescales~\citep{Moore:1998, Boselli:2006}. This general expectation contrasts with the observational evidence that a fraction of massive compact galaxies are also found outside rich clusters, in groups or lower-density environments~\citep[e.g.][]{Buitrago:2018, Tortora:2020, Scognamiglio:24}.

G\,79071 provides a clear example of this latter class. Our spatially resolved stellar population analysis shows that the galaxy is dominated by a younger stellar population (with respect to ETGs of similar mass), with luminosity- and mass-weighted ages of $\sim3$~Gyr and no evidence for a significant old component. This places G\,79071 in the broader family of massive compact systems spanning a range of mean ages, and indicates that compactness can also be established at relatively later epochs. In this interpretation, the compact configuration of G\,79071 was formed at low redshift and the galaxy has had limited time to undergo the sequence of minor mergers that is often invoked to drive size growth in massive galaxies. This naturally explains why the system still appears ultra-compact at $\rm z\sim0.1$, despite residing outside high-density regions. In principle, compact massive galaxies formed at low redshift are expected to be very rare; { however, the abundance of intermediate-age compact systems in low-mass group environments remains poorly constrained.}

At the same time, G\,79071 shows a metallicity higher than that of normal-size ETGs of comparable velocity dispersion, implying that chemical enrichment was particularly efficient in this system, that is, a rapid build-up of metals and/or star formation from already pre-enriched gas, possibly aided by efficient metal retention in the
deep potential well. The enhanced \nafe\ relative to the average SDSS population at similar $\sigma$ further supports an enrichment history in which metallicity-dependent yields play an important role. These properties are consistent with a dissipational formation pathway, in which metal-rich gas is driven to high central densities and
forms a compact stellar component.

The special conditions associated with the formation of G\,79071 are also supported by our IMF analysis. We find that the IMF is bottom-heavy in the central aperture and { shows evidence for remaining bottom-heavy} out to large radii (Sect.~\ref{sec:IMF}), implying a mass-excess factor $\alpha>1$ also beyond $\sim R_e$. In models of turbulent fragmentation, the characteristic stellar mass depends on the local star-forming conditions, including density/pressure, turbulence, and the thermodynamics of the gas \citep{Hopkins:2013}. The extreme compactness of G\,79071, together with its high metallicity, is therefore qualitatively consistent with a star-forming environment in which fragmentation is favoured at low masses, leading to a bottom-heavy distribution~\citep{Chabrier:2014}.

A bottom-heavy IMF also has implications for chemical evolution, and several studies have argued that IMF variations may require a time-dependent framework in order to satisfy chemical constraints
\citep[e.g.][]{Weidner:2013, Ferreras:2015, Fontanot:18}. For G\,79071, however, the modest enhancement of alpha elements (e.g.~\mgfe$\simeq0.1$\,dex) indicates that the chemical pattern cannot be explained by a single, short, top-heavy burst, which would generally increase \mgfe . A more plausible interpretation is that the compact star-forming episode proceeded from gas that was already metal-rich and comparatively Fe-enriched, such that the observed \mgfe\ reflects the chemical state of the fuel rather than directly encoding the duration of the final compact assembly. The required pre-enrichment does not demand a substantial old stellar mass fraction within G\,79071 itself: it can arise externally (e.g. in already evolved progenitors) and be delivered via dissipational assembly, or through reaccretion of metal-rich gas previously expelled by stellar feedback. In this picture, the bottom-heavy IMF inferred from the spectra traces the local high-pressure star-forming conditions during the compact phase, while the modest \mgfe\ does not translate into a star-formation timescale in a straightforward way~\citep{delaRosa:2011}.

From the dynamical side, the mass normalisation implied by the bottom-heavy IMF is fully consistent with our Jeans anisotropic modeling. The best-fitting JAM models reproduce the observed $V_{\rm rms}$ profile and imply a modest projected dark-matter fraction within one effective radius, $f_{\rm DM}(<R_e)=0.18\pm0.07$
(Sect.~\ref{sec:masses}), in line with typical values found in local ETGs \citep[e.g.][]{Cappellari:2013a, Tortora:2012}. Moreover, the agreement between the JAM-based mass normalization and the compactness-corrected dynamical estimator of \citet{PeraltaDeArriba:2014} highlights that departures from homology are important when interpreting dynamical masses of compact systems, whereas the classical $5\,\sigma^2 R_e/G$ estimator can
systematically underestimate $M_{\rm dyn}$ and even become physically inconsistent.

We emphasize that the present study provides particularly strong evidence for the relevance of non-homology in compact stellar systems, because it yields mutually consistent constraints from both stellar-population modeling and radially resolved dynamical analyses. This point is especially important for studies of high-redshift compact massive galaxies \citep{Slob:2025}, for which several works have reported that classical dynamical estimators can
return masses that are too low compared to the inferred stellar masses (e.g. \citealt{Saracco:2020, Kriek:2024}).

Finally, G\,79071 is relevant in a broader context because it shows that { some massive} compact ETGs with a bottom-heavy IMF can also form and persist at relatively low redshift and outside of high-density environments. This extends the phenomenology of IMF variations inferred in the central regions of massive ETGs and in compact systems such as NGC\,1277 \citep[e.g.][]{NMN:15c} beyond the earliest phases of galaxy evolution, supporting the idea that the drivers of IMF variations are linked to local star-formation physics rather than being restricted to high-redshift formation.

\section{Summary}
\label{sec:summary}

In this paper we have presented a spatially resolved stellar population and dynamical analysis for G\,79071, an
ultra-compact ($R_{\rm e}<2$\,kpc), massive ($M_\star\sim10^{11}\,M_\odot$) ETG at $z\sim0.13$, residing  in a
{ low-mass group} environment, based on deep VLT/X-Shooter long-slit spectroscopy. Our main results can be summarized as
follows:
\begin{enumerate}
\item We measured stellar kinematics along the slit out to $\sim4\,R_{\rm e}$. { G\,79071 shows significant rotation, consistent with a fast-rotator-like kinematic structure, and has a
strongly declining velocity-dispersion profile.}

\item From a combination of full spectral fitting, full-index fitting, and index fitting (using both E-MILES and
CvD18 models), we find that G\,79071 is dominated by an intermediate-age stellar population. The age profile is consistent
with being flat, with luminosity- and mass-weighted ages of $\sim3$--4\,Gyr and no evidence for a significant old
($\gtrsim5$\,Gyr) component.

\item The metallicity is supersolar in the central regions and decreases with radius, with a logarithmic gradient
typical of ETGs. Most abundance ratios show weak radial trends and are broadly consistent with those of normal SDSS
ETGs at similar velocity dispersion, while \nafe\ is significantly enhanced.

\item The stellar IMF is bottom-heavy in the central aperture and { shows evidence for remaining bottom-heavy} out to $\sim2\,R_{\rm e}$.
The corresponding mass-excess factor is $\alpha\gtrsim2$, implying stellar M/L ratios larger than those
expected for a Kroupa-like IMF.

\item Jeans anisotropic modelling (JAM) with an NFW halo reproduces the observed $V_{\rm rms}$ profile and yields a
modest projected dark-matter fraction within one effective { radius, $f_{\rm DM}(<R_{\rm e})=0.18\pm0.07$, }
consistent with typical values for local ETGs. The JAM mass normalization is in very good agreement with the IMF-based
stellar masses and with compactness-corrected (non-homologous) dynamical mass estimators, while the classical
$5\,\sigma^2R_e/G$ estimator underestimates $M_{\rm dyn}$ and can become physically inconsistent for compact systems.
\end{enumerate}

Overall, G\,79071 shows that { some massive compact ETGs} with a bottom-heavy IMF can also form and persist at relatively low redshift and outside high-density environments, suggesting that IMF variations are primarily regulated by the local physics of star formation rather than by the cosmic epoch at which a galaxy forms.  The combined stellar-population and radially resolved dynamical constraints also emphasize the importance of accounting for non-homology when interpreting dynamical masses of compact galaxies, a point that is particularly relevant for studies of compact massive systems at high redshift. Our results further suggest that environment may influence the formation and survival of massive compact galaxies, and that multiple formation channels are likely at play. Larger, well-controlled samples will be required to quantify these channels through robust estimates of comoving number densities as a function of environment.

\begin{acknowledgements}
  F.L.B. acknowledges support from the INAF minigrant 1.05.23.04.01.
F.B. acknowledges support from the GEELSBE2 project with reference PID2023-150393NB-I00 funded by MCIU/AEI/10.13039/501100011033 and the FSE+, and also the Consolidación Investigadora IGADLE project with reference CNS2024-154572.
F.B. gratefully acknowledges financial support of the Department of Education, Junta de Castilla y Le\'on and FEDER Funds (Reference: CLU-2023-1-05).

\end{acknowledgements}

\begin{appendix} 

\section{Comparison of SDSS and X-Shooter spectra}
\label{app:sdssspec}

\begin{figure}
\begin{center}
 \leavevmode
 \includegraphics[width=8cm]{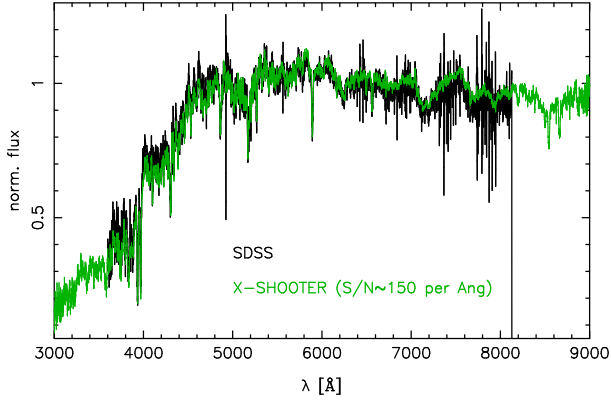}
\end{center}
 \caption{
 Comparison between the SDSS spectrum of G\,79071 (black) and the X-Shooter spectrum extracted by mimicking the
SDSS 3\arcsec-diameter fiber aperture. The two spectra show excellent agreement over the overlapping wavelength
range, with the X-Shooter spectrum reaching a significantly higher S/N ratio.
 }
   \label{fig:sdssspec}
\end{figure}

Figure~\ref{fig:sdssspec} compares the SDSS spectrum of G\,79071 (black) with the VLT/X-Shooter spectrum (green), which covers a substantially wider wavelength range. The X-Shooter spectrum was extracted by mimicking the SDSS fibre aperture, that is, by averaging the spectra along the X-Shooter slit within the SDSS fibre radius
($1.5\arcsec$). The average was computed by weighting each radial point by the flux enclosed in the corresponding elliptical annulus at that radius. Overall, the SDSS and X-Shooter spectra show excellent agreement over their common spectral range. The higher quality of the X-Shooter data is also evident, with a significantly higher
S/N ratio ($\sim150$ per \AA) and no prominent sky-subtraction residuals, which are instead visible in the SDSS spectrum.

\section{X-Shooter spectral fits for G\,79071}
\label{app:fits}

\begin{figure}
\begin{center}
 \leavevmode
 \includegraphics[width=8cm]{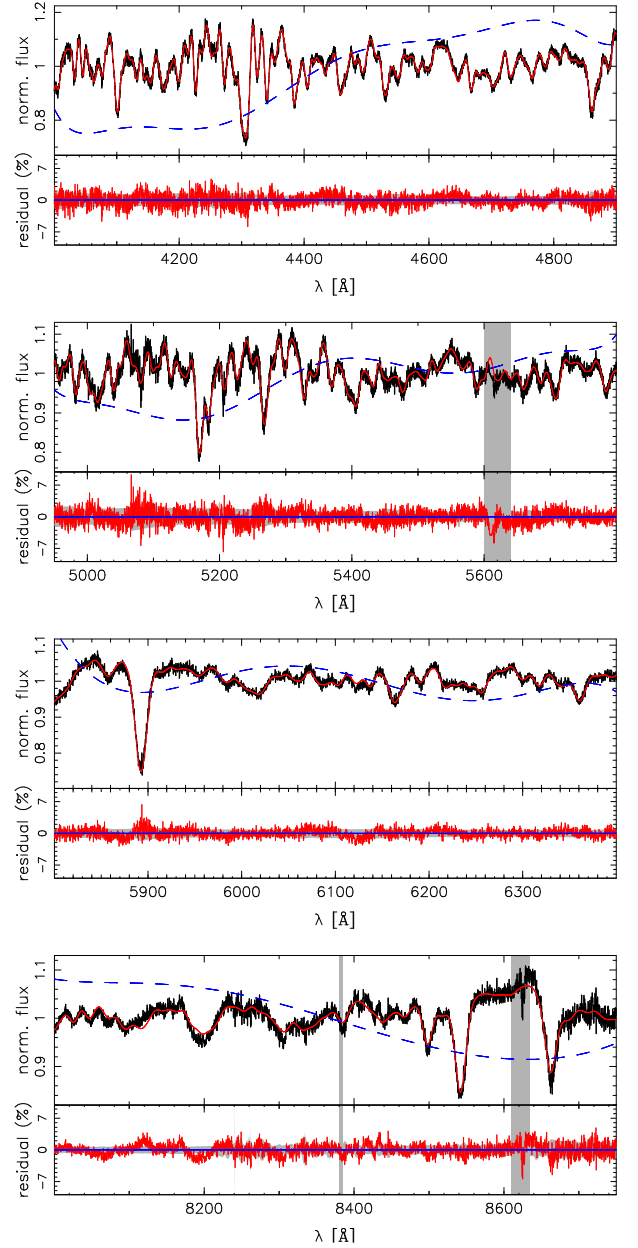}
\end{center}
 \caption{
Example of full-spectral fitting (FSF) for one of the G\,79071 spectra. The four panels correspond to the spectral ranges defined in Sect.~\ref{sec:methods}. In each panel, the upper subplot shows the observed spectrum (black) and the best-fitting model (red), both normalized by the best-fitting multiplicative polynomial (blue dashed curves), while the lower subplot shows the relative residuals (observed minus model) and the $\pm$1-sigma uncertainties (gray shaded regions). 
 }
   \label{fig:fsfexample}
\end{figure}

\begin{figure*}
\begin{center}
 \leavevmode
 \includegraphics[width=18cm]{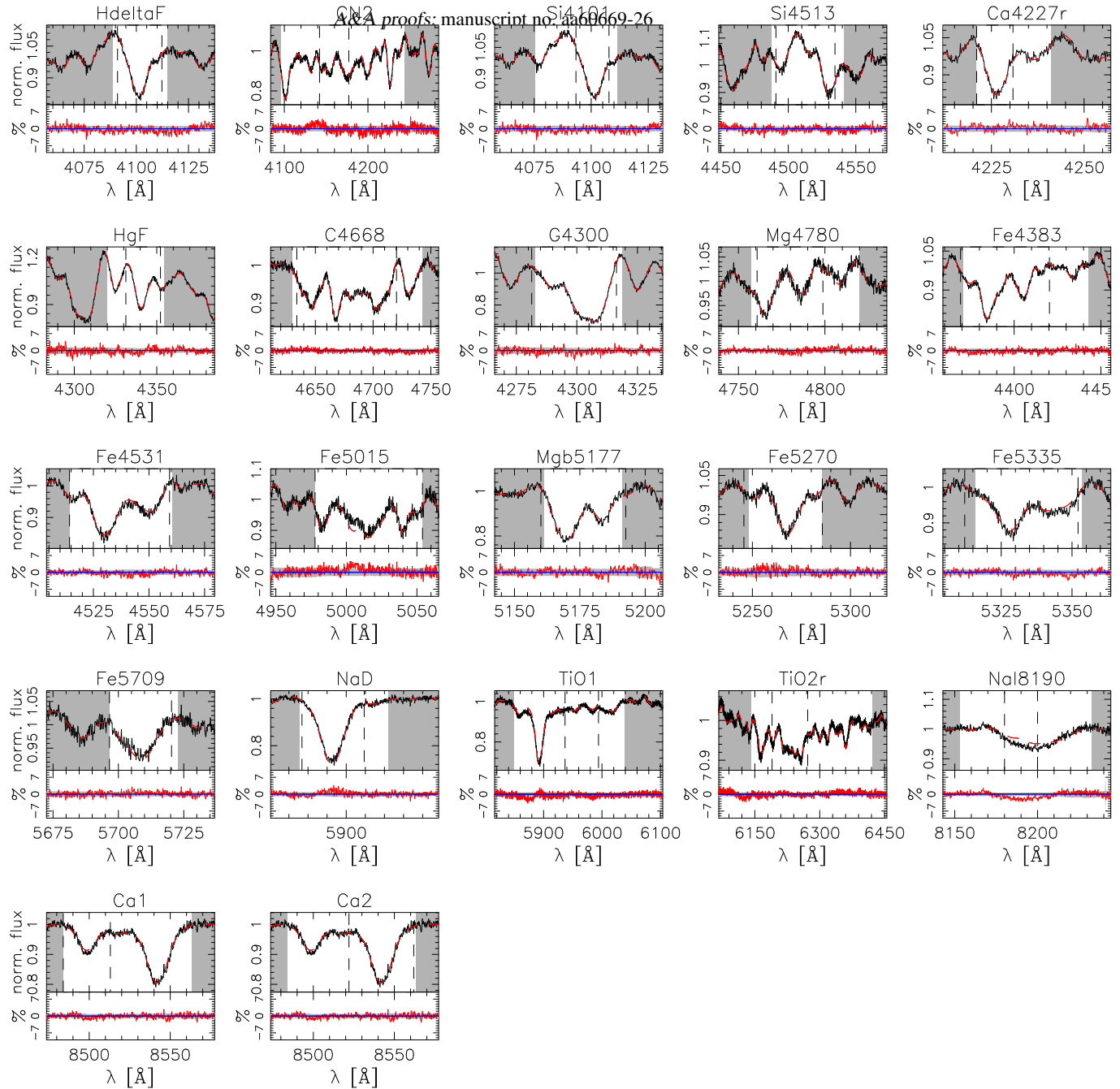}
\end{center}
 \caption{Same as Fig.~\ref{fig:fsfexample} but for full-index fitting. Different panels correspond to the spectral indices defined in Sect.~\ref{sec:methods}. In the upper subplots, gray shaded regions mark the index pseudocontinua, while vertical dashed lines define the index passbands.   
 }
   \label{fig:fifexample}
\end{figure*}

Figures~\ref{fig:fsfexample} and~\ref{fig:fifexample} show { the quality of } the full
spectral fitting and full-index fitting, respectively (Sect.~\ref{sec:methods}), for the central spectrum of
G\,79071. The black curves show the observed spectrum, while the red curves are the best-fitting two-SSP models
obtained with the E-MILES BaSTI SSPs assuming a Kroupa-like IMF, that is, a low-mass tapered (bimodal)
distribution with slope \gammab$=1.3$. For the full spectral fitting case (Fig.~\ref{fig:fsfexample}), wavelength
regions affected by reduction artefacts were masked out, as indicated by the vertical grey shaded bands.

Overall, the fits are of excellent quality, with typical rms residuals below $\sim1\%$. The largest residuals are
found around the \naii\ feature, where the best-fitting models with a standard IMF significantly underestimate the
observed absorption strength. This is consistent with the IMF constraints presented in Sect.~\ref{sec:IMF}, which
{ suggest a bottom-heavy IMF} for G\,79071 at all radii.

{ As expected, the S/N of the radially binned spectra decreases at large galactocentric distance, with different S/N thresholds adopted for the different analyses performed in this work (see Tab.~\ref{tab:binning}). For illustrative purposes, Fig.~\ref{fig:fsfexampleout} shows the full-spectral fit for the outermost radial bin used in the stellar-population analysis of G,79071. The spectrum has S/N $\sim18$ per pixel, with a pixel size of $\sim0.2,\AA$, corresponding to S/N $\sim40$ per $\AA$ (see Tab.~\ref{tab:binning}). The point-to-point scatter visible in the figure should therefore be interpreted in terms of the per-pixel S/N, rather than the S/N per $\AA$.}

\begin{figure}
\begin{center}
 \leavevmode
 \includegraphics[width=8cm]{plot_sfitp_examp_out}
\end{center}
 \caption{
{ Same as Fig.~\ref{fig:fsfexample}, but for the outermost radial bin used in the stellar-population analysis of G\,79071. The spectrum has S/N$\sim18$ per pixel, corresponding to S/N$\sim40$ per $\AA$.} }
   \label{fig:fsfexampleout}
\end{figure}

{ 
\section{Testing the presence of an old population in G\,79071}
\label{app:age_z_deg}

To assess the robustness of the age constraints for G\,79071, we
investigated the impact of the age--metallicity degeneracy on the 2SSP
best-fitting results.  Fig.~\ref{fig:AGE_Z_DEG} shows the
age--metallicity likelihood contours for the innermost radial bin of
G,79071, obtained from full-spectral fitting. The individual points
correspond to different spectral realizations, in which the flux
values are perturbed according to their uncertainties.  Results for
different IMF parametrizations are shown together.  Different colours
correspond to the different stellar population models, following the
same colour coding as in Fig.~\ref{fig:agez}.  Solid and dashed
contours mark the 1- and 2-$\sigma$ confidence levels,
respectively. The contours are well localized in the age--metallicity
plane for all stellar population models, and do not extend toward
significantly older ages.  In particular, no acceptable solution is
found with a mass-weighted age significantly larger than 5~Gyr. This
provides additional support for our conclusion that G\,79071 hosts a
genuinely young/intermediate-age stellar population.

As mentioned in Sec.~\ref{sec:sppars}, we also tested for the presence
of an old population by adding a third SSP component. In these tests,
the third component is forced to have an age older than 5~Gyr. The
fits were repeated using different initial conditions for the age and
metallicity of this component, sampling four age values from 5 to
13~Gyr and four metallicity values from $[Z/H] = -0.6$ to
$+0.2$~dex. This is done to verify that the result is not affected by
the choice of starting point in the age--metallicity plane.  For all
adopted stellar population models and IMF parametrizations, the
best-fitting contribution of the old component remained below 10 per
cent, further supporting our conclusion that G\,79071 does not host a
significant old stellar population.

\begin{figure}
 \begin{center}
\leavevmode
    \includegraphics[width=8cm]{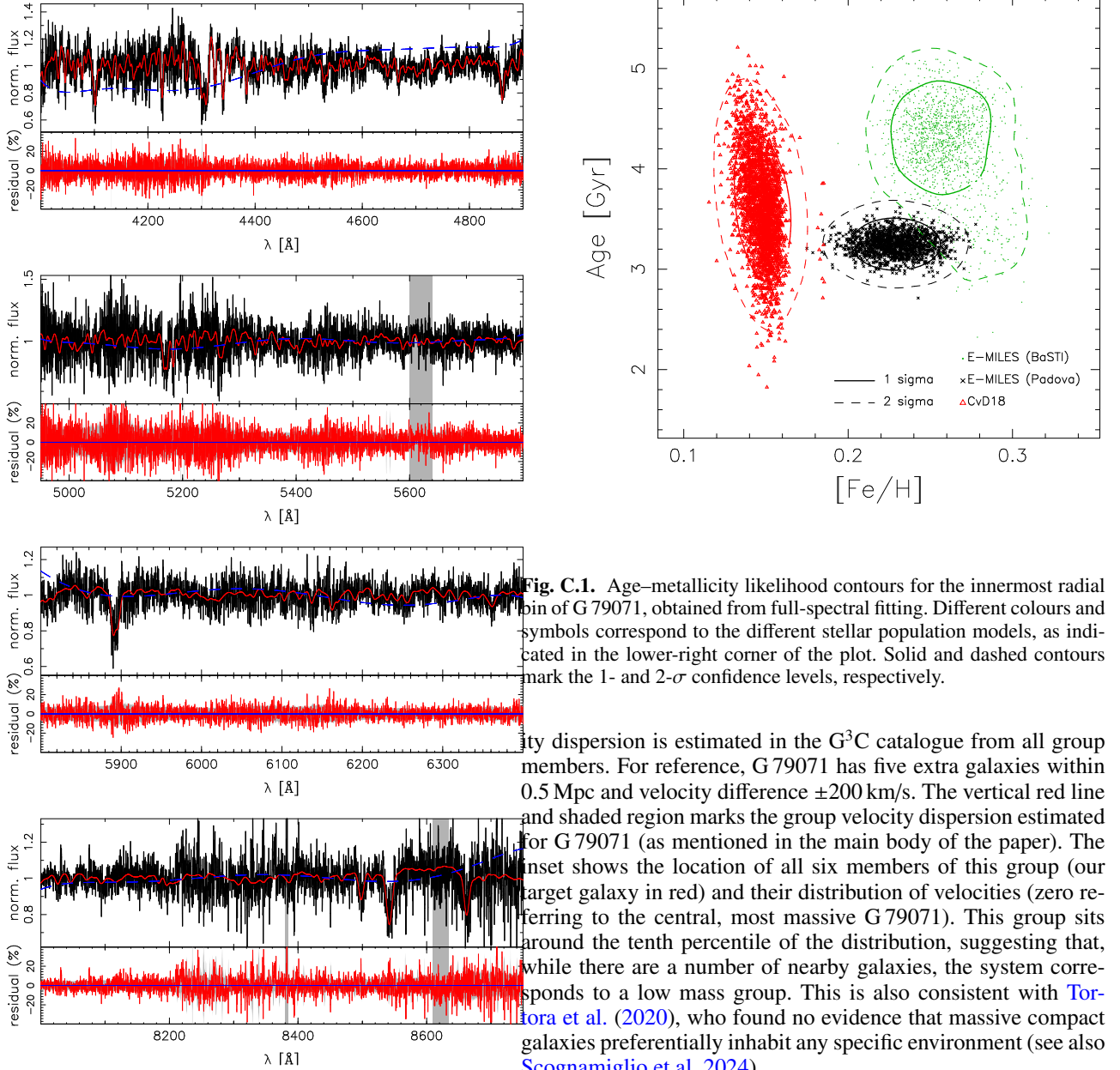}
 \end{center}
 \caption{  Age--metallicity  likelihood  contours for  the  innermost
   radial    bin    of    G\,79071,   obtained    from    full-spectral
   fitting. Different colours and  symbols correspond to the different
   stellar population  models, as indicated in  the lower-right corner
   of the plot.  Solid and dashed contours mark the  1- and 2-$\sigma$
   confidence levels, respectively. }
    \label{fig:AGE_Z_DEG}
\end{figure}
}

\section{The environment of G\,79071}
\label{app:enviro}

\begin{figure}
\begin{center}
 \leavevmode
 \includegraphics[width=8cm]{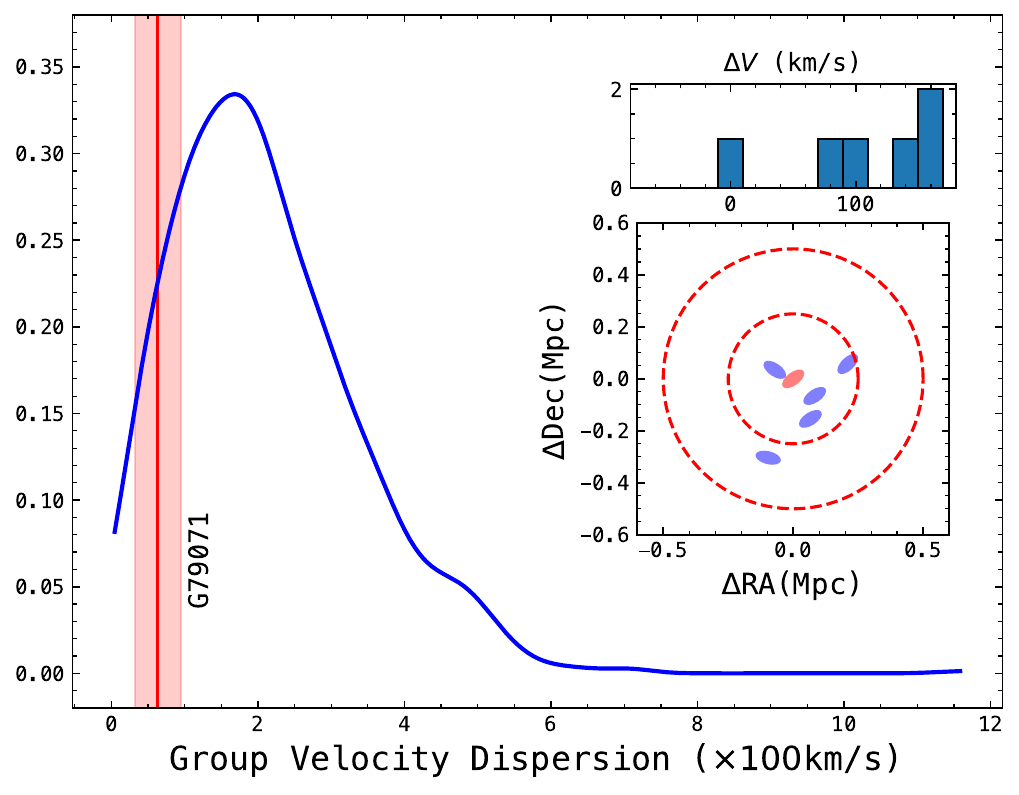}
\end{center}
 \caption{ Distribution of group velocity dispersion of GAMA galaxies from G$^3$C \citep{Robotham:11}, restricted to a similar mass and redshift as our target galaxy, whose value and uncertainty are shown with a red line and shaded region. The inset shows the projected position of all six members of the group identified for G\,79071, with dashed circles at 0.25 and 0.50\,Mpc. At the top of the inset, the distribution of velocities is shown, reflecting the clear low-mass group environment of our target. See text for details.
 }
   \label{figapp:enviro}
\end{figure}

{ We make use of the latest version of the Group Catalogue of GAMA galaxies from \citet[][G$^3$Cv10]{Robotham:11}. This catalogue produces a set of over 26,000 groups
by adopting a Friends-of-Friends algorithm that takes into account projected distance and line of sight velocity to determine simple properties of these groups. The advantage of this GAMA-based catalogue is the higher 
completeness of the survey with respect to the SDSS observations on which the survey is based \citep{Driver:11}. At the redshift of our target galaxy, GAMA is a highly complete spectroscopic survey down to  a stellar mass $\log M_{_\star}/M_\odot \sim 9.5$ \citep{Ned:11}. Fig.~\ref{figapp:enviro} shows the distribution of group velocity dispersion for a subset of GAMA galaxies restricted to have the same stellar mass as G\,79071 (within 0.3~dex in $\log M_{_\star}/M_\odot$) and similar redshift (within $\Delta z\sim 0.05$). This velocity dispersion is estimated in the G$^3$C catalogue from all group members. For reference, G\,79071 has five extra galaxies within 0.5\,Mpc and velocity difference $\pm 200$\,km/s. The vertical red line and shaded region marks the group velocity dispersion estimated for G\,79071 (as mentioned in the main body of the paper). The inset shows the location of all six members of this group (our target galaxy in red) and their distribution of velocities (zero referring to the central, most massive G\,79071). This group sits around the tenth percentile of the distribution, suggesting that, while there are a number of nearby galaxies, the system corresponds to a low mass group. This is also consistent with \citet{Tortora:2020}, who found no evidence that massive compact galaxies preferentially inhabit any specific environment (see also \citealt{Scognamiglio:24}).}

\begin{table*}
\caption{{Comparison between G\,79071 and the nearby relic galaxy NGC\,1277}}
\label{tab:ngc1277_vs_g79071}
\centering
\begin{tabular}{lcc}
\hline\hline
 & G\,79071 & NGC\,1277 \\
\hline
\multicolumn{3}{c}{Structural parameters} \\
\hline
$R_{\rm e}$ (g band) [kpc] & $2.37 \pm 0.01$ & $1.32 \pm 0.04$ \\
$R_{\rm e}$ (z band) [kpc] & $1.69 \pm 0.09$ & $1.27 \pm 0.01$ \\
Sérsic index $n$ (g band) & $2.13 \pm 0.05$ & $2.56 \pm 0.08$ \\
Sérsic index $n$ (z band) & $3.57 \pm 0.24$ & $2.64 \pm 0.15$ \\
Axis ratio $b/a$ (g band) & $0.70 \pm 0.10$ & $0.55 \pm 0.01$ \\
Axis ratio $b/a$ (z band) & $0.68 \pm 0.10$ & $0.54 \pm 0.01$ \\
\hline
\multicolumn{3}{c}{Kinematics} \\
\hline
Central velocity dispersion, $\sigma_0$ [km\,s$^{-1}$] & $\sim220$ & $\sim430$ \\
Rotation velocity at $R_{\rm e}$, $V_{\rm rot}$ [km\,s$^{-1}$] & $140$ & $280$ \\
$V_{\rm rot}/\sigma_0$ & $0.64$ & $0.65$ \\
\hline
\multicolumn{3}{c}{Stellar populations} \\
\hline
Age [Gyr] & $\sim3$ & $\sim13$ \\
Central metallicity [Z/H] [dex] & $\sim 0.2$  & $\sim 0.4$ \\
Metallicity gradient [dex$^{-1}$] & $-0.20$ & $-0.40$ \\
$[\mathrm{Mg/Fe}]$ [dex] & $\sim 0.15$ & $0.35 \pm 0.15$ \\
IMF slope \gammab & $\sim 2.9$ & $\sim2.8$ \\
\hline
\multicolumn{3}{c}{Dark matter and environment} \\
\hline
$f_{\rm DM}(<R_{\rm e})$ & $0.18\pm0.07$ & $<0.005$ \\
Local overdensity velocity dispersion,
$\sigma_{\rm env}$ [km\,s$^{-1}$]
& $63 \pm 31$ & $1040^{+34}_{-43}$ \\
\hline
\end{tabular}
\tablefoot{
{The structural parameters for 
G\,79071 come from \citet{Buitrago:2018}, while those of 
NGC\,1277 are derived in Buitrago et al. (in prep.).
The kinematic and stellar population properties of NGC\,1277 are from
\citet{NMN:15c} and \citet{FerreMateu:2017}.
The corresponding measurements for G\,79071 are derived in this work unless
otherwise noted. The dark matter fraction of NGC\,1277 is taken from \citet{Comeron:2023}. Environmental velocity dispersions come from \citet{Robotham:11} and \citet{Aguerri:2020} for G\,79071 and NGC\,1277, respectively.
}}
\end{table*}

\section{Comparison with NGC\,1277}
\label{app:ngc1277}

{Table~\ref{tab:ngc1277_vs_g79071} compares the main properties of G\,79071 with those of NGC\,1277, the prototypical nearby relic galaxy, located at $D=73$~Mpc in the Perseus cluster. NGC\,1277 is structurally more extreme than G\,79071: its stellar mass surface density within the effective radius is $\Sigma_{\rm NGC1277,2D}\sim4\times10^{10},M_\odot,{\rm kpc}^{-2}$, compared to $\Sigma_{\rm G79071,2D}\sim10^{10},M_\odot,{\rm kpc}^{-2}$. At the same time, the two galaxies share several qualitative similarities. Both appear visually smooth and disk-like, and their kinematics indicate substantial rotational support, consistent with a fast-rotator-like structure. Their IMF constraints are also similar, with bottom-heavy slopes that remain approximately constant out to $\sim2,R_{\rm e}$ and correspond to mass-excess factors $\alpha \gtrsim 2$ relative to a Kroupa-like IMF. The stellar-population properties are, however, more extreme in NGC\,1277, which is older, more metal-rich, and more $\alpha$-enhanced than G\,79071. Dynamically, both systems are baryon dominated in their central regions, as commonly found for nearby early-type galaxies. However, NGC\,1277 appears to be extremely dark-matter deficient within the stellar body, whereas G\,79071 is not exceptional in this respect. Their environments are also markedly different: NGC\,1277 resides in the high-mass Perseus cluster, while G\,79071 is associated with a low-mass galaxy group, with the corresponding environmental velocity dispersions differing by almost two orders of magnitude. These differences suggest that massive compact galaxies with bottom-heavy IMFs may not all follow a single evolutionary pathway. Another potentially relevant ingredient is the central supermassive black hole. In NGC\,1277, feedback from its unusually massive black hole has been proposed as a possible contributor to its quiescent evolution. However, the interplay between galaxy growth and black-hole evolution in red nuggets and relic galaxies remains poorly understood and is still the subject of ongoing debate.}

\end{appendix}

\end{document}